\documentclass[sn-mathphys,Numbered]{sn-jnl}

\usepackage{graphicx}%
\usepackage{multirow}%
\usepackage{amsmath,amssymb,amsfonts}%
\usepackage{amsthm}%
\usepackage{mathrsfs}%
\usepackage[title]{appendix}%
\usepackage{xcolor}%
\usepackage{textcomp}%
\usepackage{manyfoot}%
\usepackage{booktabs}%
\usepackage{algorithm}%
\usepackage{algorithmicx}%
\usepackage{algpseudocode}%
\usepackage{listings}%
\usepackage{anyfontsize}
\usepackage{float}
\usepackage{pgf}
\usepackage{color, colortbl}
\usepackage{todonotes}
\usepackage{natbib}
\usepackage{multirow}
\usepackage{textgreek}
\usepackage{hyperref}
\usepackage{footnote}
\usepackage{pgfplots}
\usepackage{colortbl}
\usepackage{lineno}

\pgfplotsset{compat=1.18}

\begin{document}

\title[Article Title]{Polymer Micro-Lattice buffer structure Free Impact absorption}

\author*[1]{\fnm{Louis} \sur{Catar}}\email{louis.catar.1@ens.etsmtl.ca}

\author[2]{\fnm{Ilyass} \sur{Tabiai}}

\author[1]{\fnm{David} \sur{St-Onge}}

\affil[1]{\orgdiv{INIT Robots}, \orgname{École de Technologie Supérieure}, \orgaddress{\street{1100 rue Notre-Dame O.}, \city{Montréal}, \postcode{H3C 1K3}, \state{QC}, \country{Canada}}}

\affil[2]{\orgdiv{LIPEC}, \orgname{École de Technologie Supérieure}, \orgaddress{\street{1100 rue Notre-Dame O.}, \city{Montréal}, \postcode{H3C 1K3}, \state{QC}, \country{Canada}}}


\abstract{The uncrewed aerial systems industry is rapidly expanding due to advancements in smaller electronics, smarter sensors, advanced flight controllers, and embedded perception modules leveraging artificial intelligence. These technological progress have opened new indoor applications for UAS, including warehouse inventory management, security inspections of public spaces and facilities, and underground exploration. Despite the innovative designs from UAS manufacturers, there are no existing standards to ensure UAS and human safety in these environments. This study explores developing and evaluating micro-lattice structures for impact resistance in lightweight UAS. We examine patch designs using Face-Centered Cubic (FCC), Diamond (D), Kelvin (K), and Gyroid (GY) patterns and detail the processes for creating samples for impact and compression tests, including manufacturing and testing protocols.

Our evaluation includes compression and impact tests to assess structural behavior, revealing the influence of geometry, compactness, and material properties. Diamond and Kelvin patterns were particularly effective in load distribution and energy absorption over the compression tests. Impact tests demonstrated significant differences in response between flexible and rigid materials, with flexible patches exhibiting superior energy dissipation and structural integrity under dynamic loading.

The study provides a detailed analysis of specific energy absorption (SEA) and efficiency, offering insights into optimal micro-lattice structure designs for impact resistance in lightweight UAS applications. 


}




\maketitle

\section{Introduction}

The uncrewed aerial systems (UAS) industry is rapidly expanding due to advancements in smaller electronics, smarter sensors, advanced flight controllers, and embedded perception modules leveraging artificial intelligence. These technological progresses have opened new indoor applications for UAS, including warehouse inventory management, security inspections of public spaces and facilities, and underground exploration. Despite the innovative designs from UAS manufacturers, there are no existing standards to ensure human and UAS safety in these environments. As these aerial vehicles are integrated into more tasks, they operate closer to operators and users, calling for new safety measures for safe human-UAS interaction. Additionally, several recent applications target indoor cluttered spaces, requiring protection of their valuable payloads from impacts~\cite{noauthor_elios_nodate}.

Polymer materials are widely used across various industrial sectors due to their diverse manufacturing processes and the extensive range of properties they can achieve. However, their low mechanical strength often limits their use in aerospace applications. Polymer micro-lattice structures can address this issue by providing enhanced specific mechanical properties through their intricate cell patterns. Micro-lattices are structures composed of periodic cells, granting variable mechanical properties compared to bulk material~\cite{zheng_ultralight_2014}. The mechanical properties of a micro-lattice depend on the interplay between the material properties, the geometry of the unit cell, the macroscopic structure of the part, the density of the structure (cells per volume), and the manufacturing technique~\cite{evans_multifunctionality_1998}. These different levels of complexity make each use case unique. 

Micro-lattice structures have a wide range of potential applications in various fields such as aerospace, biomedical engineering, and energy storage~\cite{yeo_structurally_2019}. In aerospace, micro-lattice structures can be used to develop lightweight structures with high specific strength and stiffness, which can improve the efficiency and performance of aircraft and spacecraft. In biomedical engineering, these structures can be used to create scaffolds for tissue engineering or drug delivery systems because their high surface area and porosity enable efficient transport of nutrients and drugs~\cite{fina_3d_2018}. In energy storage, micro-lattice structures can be used as electrodes in batteries and supercapacitors, as their high surface area and porosity enable rapid charge and discharge rates~\cite{xiong_advanced_2015}. Despite their potential, experimental studies are still necessary to better understand the failure mechanisms of micro-lattices~\cite{wu_additively_2023}. While our study focuses on small aerial vehicle impact protection, the findings can be extended to these other domains.

Despite their potential, micro-lattices are not yet common in commercial products due to the complexity of their manufacturing. The advent of polymer additive manufacturing (AM) offers the opportunity to build complex geometric structures at low costs. However, not all AM techniques meet the high-resolution and support-free requirements essential for micro-lattice production~\cite{nazir_state---art_2019}. One viable option is Masked Stereolithography Apparatus (MSLA) technology, which solidifies photosensitive resin with a high level of detail on affordable machines with a printing resolution of around 20 to 50 µm.

A wide variety of polymer materials have been proposed for impact protection. In the aerospace sector, micro-lattice sandwich panels are considered an alternative to 2D-honeycomb structures for shielding sensitive equipment~\cite{mines_drop_2013}. For single-impact applications, expanded polystyrene foam (EPS), and rigid polyurethane foam (PU) are commonly used in the transportation and aerospace industries for their potential to reduce weight while enhancing safety~\cite{blazy_comportement_2003}. For example, EPS is often used in motorcycle and bicycle helmets, but these helmets must be discarded after a single impact. For multiple-impact applications, expanded polypropylene foam (EPP), PU foam, and vinyl-nitrile foam (VN) are recommended. In combat helmets and American football helmets, shock attenuation micro-lattice structures with multi-impact capability are employed to maximize protection against repeated impacts~\cite{clough_elastomeric_2019}. Various other industries are investigating polymer structures capable of withstanding significant impacts while maintaining their functional integrity in applications such as shoe soles, airless tires, and biomorphic prostheses~\cite{nazir_state---art_2019}. The application's requirements also shape the way these structures are tested: impact tests are categorized into four families based on the speed of the impact, or application: low-velocity test (below 10 m/s), intermediate velocity test (10 m/s to 50 m/s), ballistic velocity (50 m/s to 1000 m/s) and high-velocity impact (1000 m/s to 5000 m/s)~\cite{ismail_low_2019}. Despite the growing interest across these domains, there remains a limited understanding of the mechanical characteristics and variability among different micro-lattice patterns.

As for the cell geometry, they can be grouped into three families: micro-lattices composed of struts, triply periodic surfaces (TPMS), and shell micro-lattices. Up to 2016~\cite{sakshikokil-shah_recent_2021}, BCC patterns were the most studied because they are easy to manufacture using a wide range of processes. Almahri et al.~\cite{almahri_evaluation_2021} compared five patterns of TPMS metallic surface micro-lattices. TPMS limits macro defects thanks to the continuity of the structure, which limits deformations during fabrication.

Only a handful of works have been published on the impact performance of micro-lattices, and they all focus on metallic parts. Xiao et al.~\cite{xiao_additively-manufactured_2018} studied the influence of step-wise density gradient versus a uniform structure in TPMS manufactured in Titanium and Aluminum alloys via SLM. Samples were subjected to high-velocity impacts using a Split Hopkinson bar device. The results showed a superior SEA by 28\% for gradient micro-lattices. 
Evans et al.~\cite{evans_concepts_2010} worked on micro-lattices with hollow nickel tubes to provide higher energy absorption per unit mass~\cite{xiong_advanced_2015}. With blast tests, they compared them to stochastic foams and honeycomb structures. They observed four to five times the SEA of foams and honeycomb in experimentation, which they explained with the variety of deformation modes, such as cell buckling and accordion folding. Cui et al.~\cite{cui_dynamic_2012} confirms this finding on their tetrahedral micro-lattice panels compared to honeycomb structures. They also highlighted densification regions under the impacter, followed by a transition zone.

With the availability of consumer-grade AM machines that cost a few hundred dollars, research institutions have started to push their limits~\cite{catar_additive_2022, hassan_design_2023}. These new additive manufacturing processes have unlocked a wide range of possibilities for process technique variations and the versatility of micro-lattice geometries. This finding makes the characterization of mechanical performance and the use of micro-lattices increasingly relevant.


The contributions of this paper are \textit{(1)} the study of micro-lattices for low-speed impact attenuation in realistic experimental conditions; \textit{(2)} a classification of loading cases relative to the flexibility of the material; \textit{(3)} to the best of our knowledge, the first studies on micro-lattices made by MSLA machine and at a low density (under 100~g/m\textsuperscript{3}) and \textit{(4)} our micro-lattices patches emphasizes a significant reduction in the relative density of micro-lattices while maintaining comparable effectiveness in absorbing impact energy.

\section{Micro-lattices energy absorption behavior} \label{sec:related}

Before looking into the impact mechanics and the performance of the micro-lattice patches, it is essential to provide an overview of the mechanisms involved in energy absorption specific to micro-lattices. Guillon~\cite{guillon_etude_2008} provides a comprehensive analysis of the physical phenomena of energy absorption. Polymers exhibit viscoelastic properties, reacting differently depending on the loading speed. During impacts, when polymer chains do not have sufficient time to rearrange, this results in increased material stiffness and strength~\cite{menard_dynamic_2020}. Absorption results from the plastic deformation of the structure, occurring over two distinct phases. The first phase involves the initial damage, which will stabilize to form a crushing front, often characterized by a peak in the load curves. Following this, the load may stall, with contribution from different deformation modes. The “folding mode” or “local buckling mode” is the most common for low-stiffness polymers, where energy is dissipated through localized buckling and yielding, reflected by regular oscillations in the force/displacement curve corresponding to fold formations. Rupture modes and the evolution of the propagation front can be controlled by adding triggers such as notches~\cite{guillon_etude_2008}. Depending on the material's rigidity, we can induce micro-lattice destruction or a consolidation phase where the micro-lattice folds, observable as a force peak~\cite{clough_elastomeric_2019}. To quantify performance, Ramakrishna et al.~\cite{ramakrishna_energy_1995} introduced the Specific Energy Absorption (SEA), defined as the energy absorption per unit mass, and derived as follows:
\begin{equation}
    SEA = \frac{E}{m} = \frac{F_{avg}}{A \rho}
    \label{eq:SEA}
\end{equation}
With $E$ as the absorption energy, $m$ as the mass, $F_{avg}$ as the average load, $A$ as the cross-section, and $\rho$ as the density.

Several studies have attempted to address SEA characterization through simulation. Finite element models generally represent experimental stiffness values accurately, but often diverge in terms of the loading history of the structures~\cite{gumruk_compressive_2013}. This divergence can be caused by structural instabilities~\cite{xiong_advanced_2015} or manufacturing imperfections~\cite{wu_additively_2023}, directly impacting the realism of the simulations. Rathbun et al.~\cite{rathbun_performance_2006} found that dynamic SEA values are greater than those observed in quasi-static conditions, further complicating the simulation of load cases.
Zhang et al.~\cite{zhang_energy_2021} achieved experimental results consistent with their theoretical model predictions and finite element analysis, with errors of less than 22\%. They focused on compression tests over BCC/FCC-type structures, noting that the addition of vertical struts significantly increases the specific strength and SEA, whereas diagonal struts have a lesser effect. Realistic finite element studies of micro-lattice structures become extremely complex when the size of the structure necessitates numerous mesh elements~\cite{smith_finite_2013}. Nevertheless, basic properties can be extracted by studying a single micro-lattice cell. Finite element models can provide good approximation of the maximum stress values during compression or impact but cannot accurately represent local deformations or local damage to the micro-lattice~\cite{feng_mechanical_2022}.

Li et al.~\cite{li_compressive_2006}~introduced a metric of energy efficiency $\eta$. It accounts for the efficiency of absorption of a cellular architecture, foam, or micro-lattice. This puts the SEA into perspective respecting the theoretical maximum energy that the structure would be capable of absorbing. 

It is defined as:
\begin{equation}\label{eq:eta}
    \eta = \frac{1}{\sigma_p} \ \int_{0}^{\epsilon_d} \sigma (\epsilon) \,d \epsilon \
\end{equation}

$\sigma_p$ is the maximum stress before densification. In this equation, the ratio $\eta$ is calculated between the area under the stress-strain curve of the structure and that of an ideal material, keeping a maximum stress $\sigma_p$ and a maximum strain of 1. Based on the data from the literature, the information from \cite{schaedler_designing_2014} is particularly interesting because they also conducted quasi-static compression tests (1 mm/min) and impact tests (5 m/s). We will compare it with our values in Section~\ref{sec:SEA-eta}.

\section{Methods and materials}\label{sec:metho}

To contextualize the results of our extensive test campaign, we provide a detailed account of the design of the test patches (specimens), their manufacturing processes, and the protocols used for compression and impact testing.

\subsection{Patch design}

The patch specimen was designed as a square with 45 mm sides and a thickness of 15 mm, which limits each patch's weight to 10 g, including 2 g of micro-lattice. The 15 mm thickness can consist of a single layer of lattice cells, 2 layers or at most 3 layers (5mm each).

\begin{figure}[H]
\begin {center}
\includegraphics[width=0.5\textwidth]{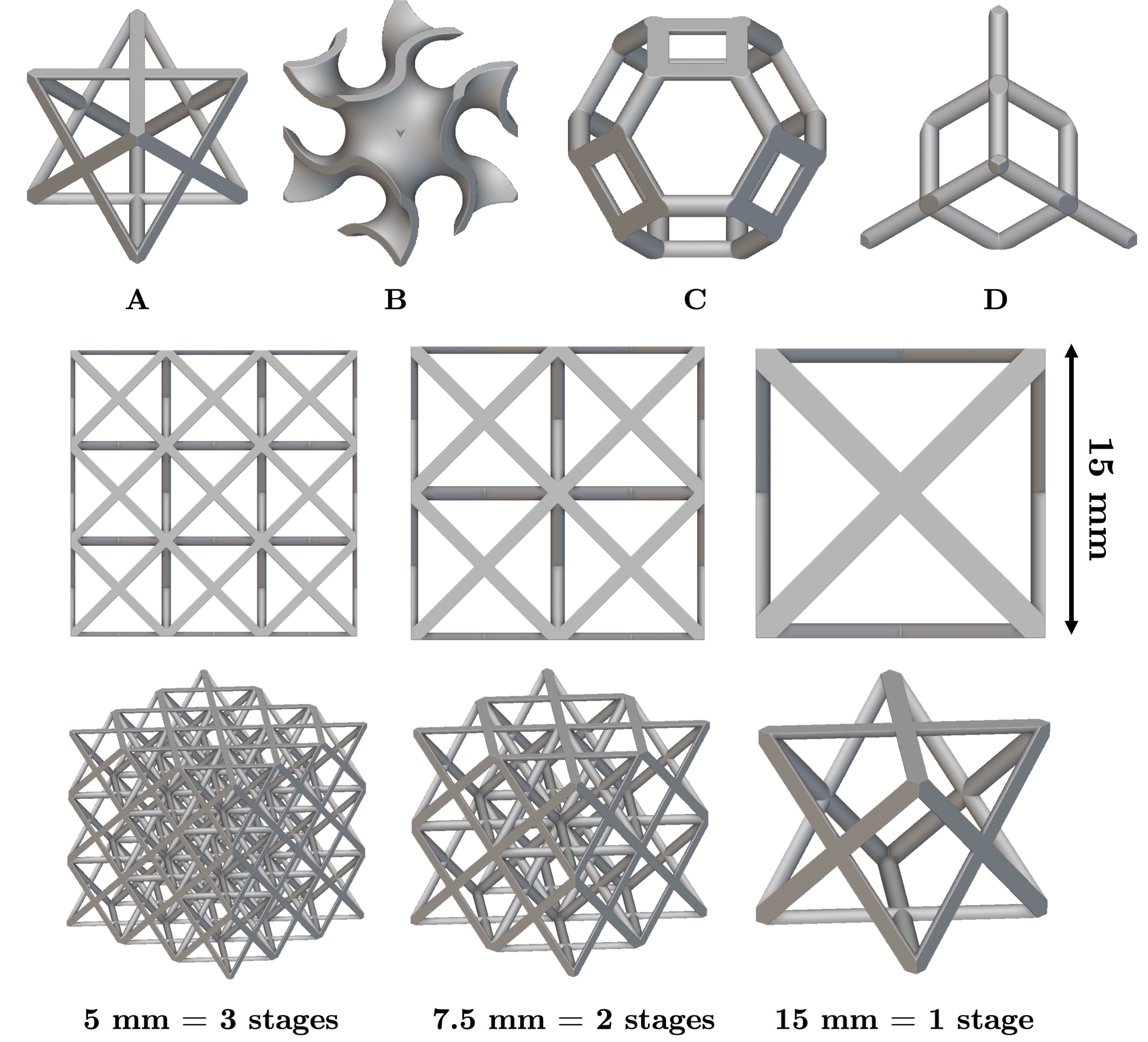}
\caption{Micro-lattices pattern selection: Face Centered Cubic—FCC (\textbf{A}), Gyroid—GY (\textbf{B}), Kelvin—K (\textbf{C}), and Diamond—D (\textbf{D}) and example of the three levels of compactness to fill the 15 × 15 × 15 mm\textsuperscript{3} volume with the FCC pattern. The number of micro-lattice stages changes, but the diameters of the struts are adapted to have the same relative mass. The patch samples have a total volume of 45 × 45 × 15 mm\textsuperscript{3}. }
\label{fig:micro-lattice}
\end {center}
\end{figure}

We selected four micro-lattice patterns: Face Centered Cubic, Diamond, Kelvin, and Gyroid. We focused on those providing mechanical properties relevant to ultra-light aerospace structures plus three criteria: \textit{(1)} providing a reference base comparable with previous studies, \textit{(2)} aiming for a wide diversity in typologies, and \textit{(3)} ensuring a self-supporting cell geometry for all manufacturing processes~\cite{catar_micro-tensile_2022}. The FCC pattern (Fig.~\ref{fig:micro-lattice}-A) ensures stability during the manufacturing process and has been extensively studied. The FCC nodes serve as junctions for more struts, thereby reducing stress concentrations. Secondly, the diamond pattern (Fig.~\ref{fig:micro-lattice}-B) offers geometric homogeneity and lightweightness. Its geometry is similar to that of a honeycomb strut-based 3D structure. Then, we add the Kelvin pattern (Fig.~\ref{fig:micro-lattice}-C), which has large internal cavities. The planar and inclined strut orientations of the Kelvin pattern provide interesting structural rigidity for dissipating energy during an impact~\cite{saremian_experimental_2021}. This lattice features longitudinal struts in the loading direction, potentially leading to behavioral differences and visible instabilities during testing, as shown by Xiong et al.~\cite{xiong_advanced_2015}. The gyroid pattern (Fig.~\ref{fig:micro-lattice}-D), known for its high-energy absorption under compression~\cite{abueidda_mechanical_2019}, represents the family of triply periodic surfaces (TPMS). Fig.~\ref{fig:micro-lattice} shows the geometry and related information. 


In addition to comparing different micro-lattice patterns, we look into the influence of micro-lattice compactness, or density. We defined three levels of compactness: cells of 15 mm, 7.5 mm, and 5 mm, allowing for 1 to 3 layers of micro-lattice in the 15 mm patch as shown in Fig.~\ref{fig:micro-lattice}. We designed the volumes and mounting plates for all patches using Autodesk Fusion 360 software. The space allocated for micro-lattices was filled using nTopology software, which provides extensive options for creating micro-lattices with its implicit design engine. Among the different variants, the thickness of the struts or walls was adjusted to ensure that all samples had the same mass. Table~\ref{tab:thickness} presents the diameters or wall thickness values for each of the patterns and their respective compactness levels.

\begin{table}[h]
\label{tab:thickness}
\caption{Diameter of the struts or thickness of the shell of micro-lattices in millimeters to keep all patches at 2 g weight}
\begin{tabular}{|l|ccc|}
\hline
\multicolumn{1}{|c|}{}                          & \multicolumn{3}{c|}{\textit{\textbf{Compacity}}}                                                                                                                                                                                \\ \hline
\multicolumn{1}{|c|}{\textit{\textbf{Pattern}}} & \multicolumn{1}{c|}{\cellcolor[HTML]{008080}{\color[HTML]{FFFFFF} \textbf{50}}} & \multicolumn{1}{c|}{\cellcolor[HTML]{0082C8}{\color[HTML]{FFFFFF} \textbf{75}}} & \cellcolor[HTML]{F58230}{\color[HTML]{FFFFFF} \textbf{150}} \\ \hline
\textbf{$\blacksquare$ FCC}                                    & \multicolumn{1}{c|}{0.48}                                                       & \multicolumn{1}{c|}{0.73}                                                       & 1.5                                                         \\ \hline
\textbf{$\blacklozenge$ D}                                      & \multicolumn{1}{c|}{0.52}                                                       & \multicolumn{1}{c|}{0.78}                                                       & 1.56                                                        \\ \hline
\textbf{$\bigstar$ K}                                      & \multicolumn{1}{c|}{0.48}                                                       & \multicolumn{1}{c|}{0.72}                                                       & 1.46                                                        \\ \hline
\textbf{$ \bullet$ GY}                                     & \multicolumn{1}{c|}{0.275}                                                      & \multicolumn{1}{c|}{0.57}                                                       & 0.82                                                        \\ \hline
\end{tabular}
\end{table}

We also studied two different materials: a flexible resin (Tenacious\footnote{\href{https://drive.google.com/file/d/1COV6emzdbWFPRHsKVQWdIzHywZKDzrIx/view}{Datasheet Tenacious (Soft)}}) and a rigid resin (Build\footnote{\href{https://drive.google.com/file/d/1BwUNGdiv5DE_r0xTPJs-xIbd2-2avMXO/view}{Datasheet Build (Rigid)}}). Both materials are chemically similar, derived from the same urethane acrylate base, and manufactured by the same company, SirayaTech\textsuperscript{\textregistered}. It is even possible to mix them to create new material variations. The densities of the materials were measured using a hydrostatic balance with a cubic specimen at 100\% infill. The rigid material showed a density of 1.237~g/cm\textsuperscript{3} $\pm$ 0.001 while the flexible one had a density of 1.214~g/cm\textsuperscript{3} $\pm$ 0.08. The manufacturer provides values of 33 MPa for ultimate tensile stress with 8\% elongation at break for the rigid resin using ASTM D638. For the flexible resin, the values are 5 MPa for tensile stress at break and 70\% elongation at break. We conducted mechanical tensile tests following the ASTM D638 standard to validate these values. For the rigid resin, we obtained 22.7 MPa with 3\% elongation (underperformance), and for the flexible resin, we obtained 7.9 MPa with 8\% elongation (overperformance). These discrepancies from the manufacturer's data are likely due to the duration of UV exposure during printing and curing. Results can also vary depending on room humidity and the duration of alcohol exposure when cleaning the parts.

\begin{figure}[H]
\begin {center}
\includegraphics[width=0.4\textwidth]{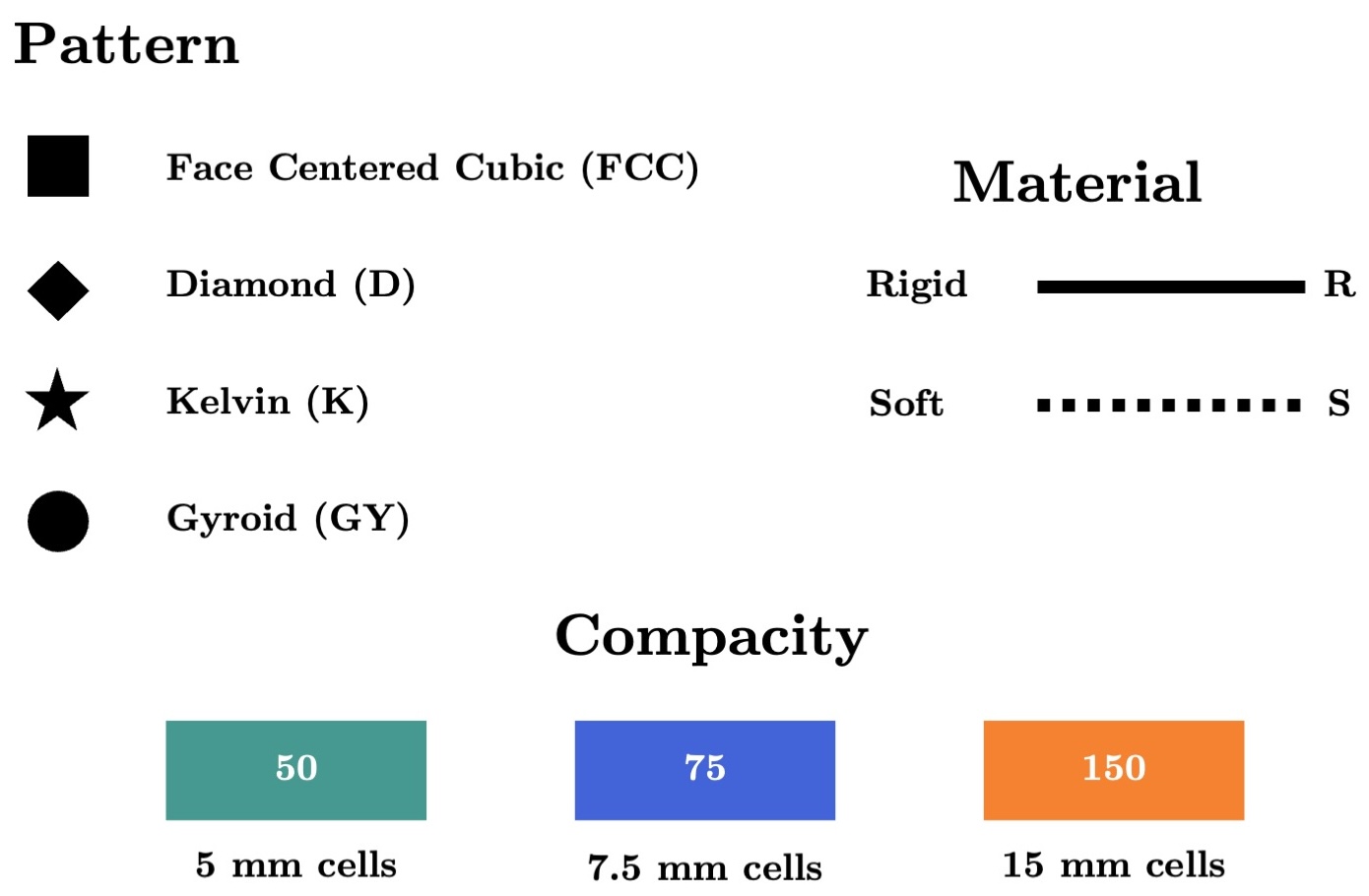}
\caption{Figure nomenclature: There are so many possible combinations that we've created this nomenclature, which serves as a standard for all the figures in this paper. }
\label{fig:nomenclature}
\end {center}
\end{figure}

Throughout the paper, we will use the following abbreviations: FCC, GY, K, and D, for the micro-lattices patterns, with the compacity 150, 75, and 50 and the resins will be abbreviated as R and S. Three samples are evaluated for each combination to assess repeatability. This brings the total number of samples to be manufactured to 144. Figure~\ref{fig:nomenclature} shows the nomenclature standard to represent the selection of patches for all graphs in this paper. 

\subsection{Patch manufacturing}\label{sec:fab}

To manufacture the micro-lattice patches, we used the MSLA AM technique. It is the best resolution-to-price ratio 3D printer technology.
The MSLA process uses an LCD screen to create a mask between the UV light and the resin tank. When cured, the polymerized resin forms small parallelepipeds (voxels) whose dimensions correspond to the pixel size in the XY plane and the layer height in the Z direction. Thus, the machine's resolution in the XY fabrication plane is directly correlated to the screen's resolution. Thanks to this voxel manufacturing, MSLA provides isotropy-like mechanical properties in the resulting parts. This isotropy is beneficial for manufacturing various micro-lattice patterns that are inherently multidirectional. The use of this technology is advantageous because MSLA machines support a wide variety of materials from different manufacturers (not a proprietary technology) and are generally composed of simple, inexpensive components. For this study, we used an \copyright~\textit{Elegoo Saturn S}, equipped with a 4K 9.1-inch LCD screen. The size of a pixel and the resolution in the XY-plane is 48 µm. The minimum layer height is 10 µm. For slicing the parts, we used the open-source Chitubox software.

Given the selected manufacturing technology, the parts must be designed to meet specific requirements. For instance, samples must be designed to avoid the need for printing supports, minimizing the risk of pattern alteration during post-processing. This approach reduces printing failures, manufacturing time, material consumption, and costs while ensuring a high printing success rate.

The patch geometry is 45 mm x 45 mm x 15 mm and must be mechanically attached to the impact bench (see Sec.~\ref{sec:bench}). Therefore, the samples require mounting holes and a structure to protect them. With a substantial contact surface with the printing bed, adhesion can be too strong for the small struts of the micro-lattices. This could result in the piece being destroyed or damaged when manually removed with a spatula at the end of the print. Conversely, if the contact surface is too small, for instance, if supported by narrow pillars, the piece may deform during polymerization. This can lead to detachment from the printing bed, halting production to remove the part, or, worse, damaging the machine if the piece becomes stuck at the bottom of the resin tank.

\begin{figure}[H]
\begin {center}
\includegraphics[width=0.4\textwidth]{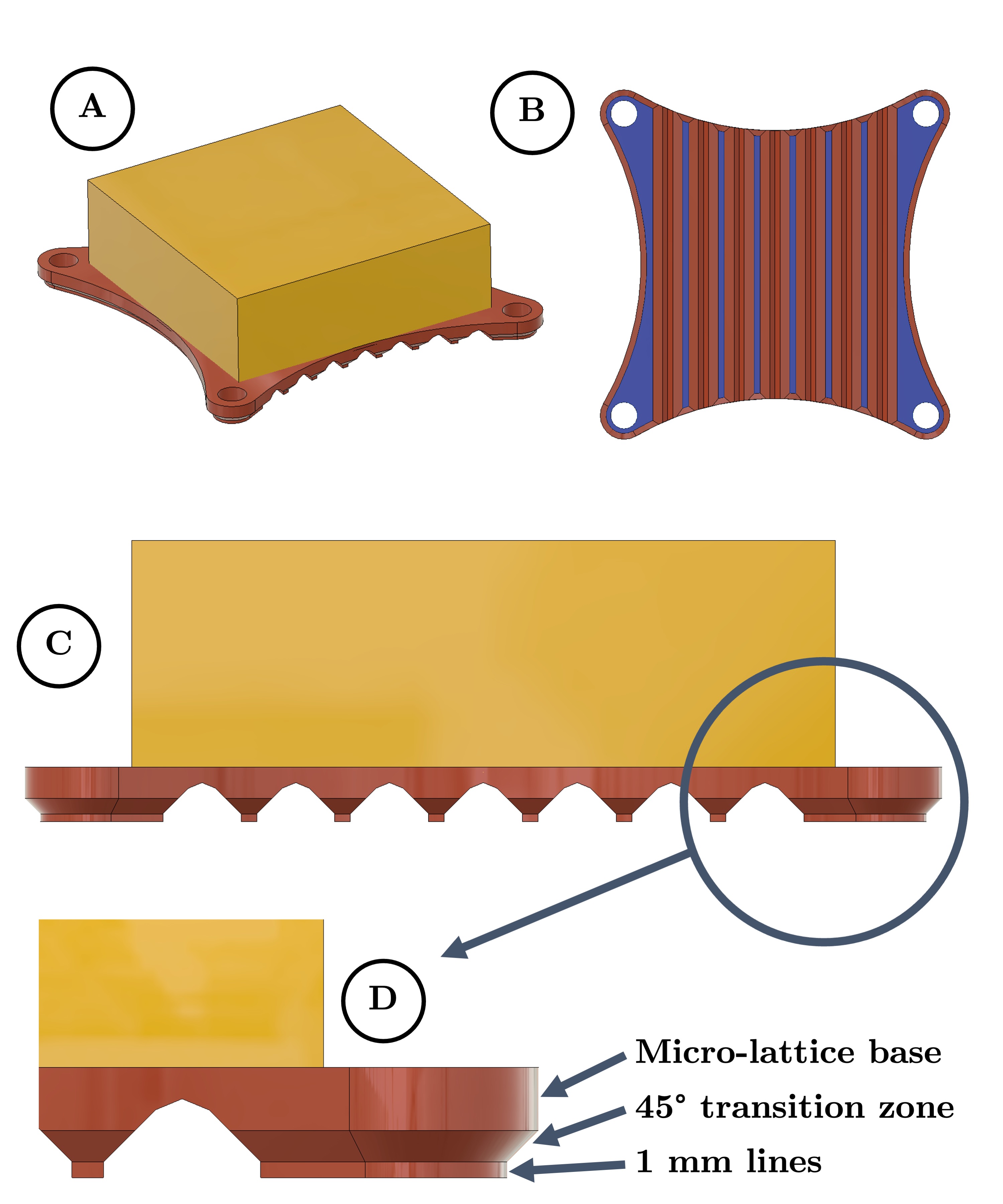}
\caption{Patch design: (\textbf{A}) Optimized sole plate design, (\textbf{B}) Furrows section with minimized surface in contact with the print bed, (\textbf{C}) a cross-section of the furrows grooves, and (\textbf{D}) the detail of the successive part layers.}
\label{fig:soleplate}
\end {center}
\end{figure}

To address these challenges, we propose an optimized sole plate for MSLA printing that provides several benefits. Firstly, it serves as a mounting interface between the micro-lattice area and various testing platforms, such as the compression bench and impact bench. Secondly, it ensures that the micro-lattices are not affected by sample manufacturing, acting as a support while allowing easy manual detachment. The optimized sole plate, shown in Fig.~\ref{fig:soleplate}, is composed of several layers, all made from the same material, each with distinct roles. 

The underneath furrows reduce the contact surface with the printer's bed, with each furrow having only a 1 mm thick plateau in contact with the printing bed. The lines then taper to create a self-supporting surface at a 45° angle to the printing direction. This area serves as a transition between the furrow zone and the platform that hosts the micro-lattice. It also provides space to manipulate the piece more easily when detaching it from the bed. Finally, the platform on which the micro-lattice is built is thick enough to be rigid, preventing any risk of flatness deformation.

The complete step-by-step manufacturing protocol can be found \href{https://git.initrobots.ca/lcatar/micro-lattices-patches-manufacturing}{here}.

\subsection{Patch compression tests}

Compression tests are useful in the preamble of impact tests because they provide detailed information on how the cells of the micro-lattice deform under a constant load.

\begin{figure}[H]
\begin {center}
\includegraphics[width=0.45\textwidth]{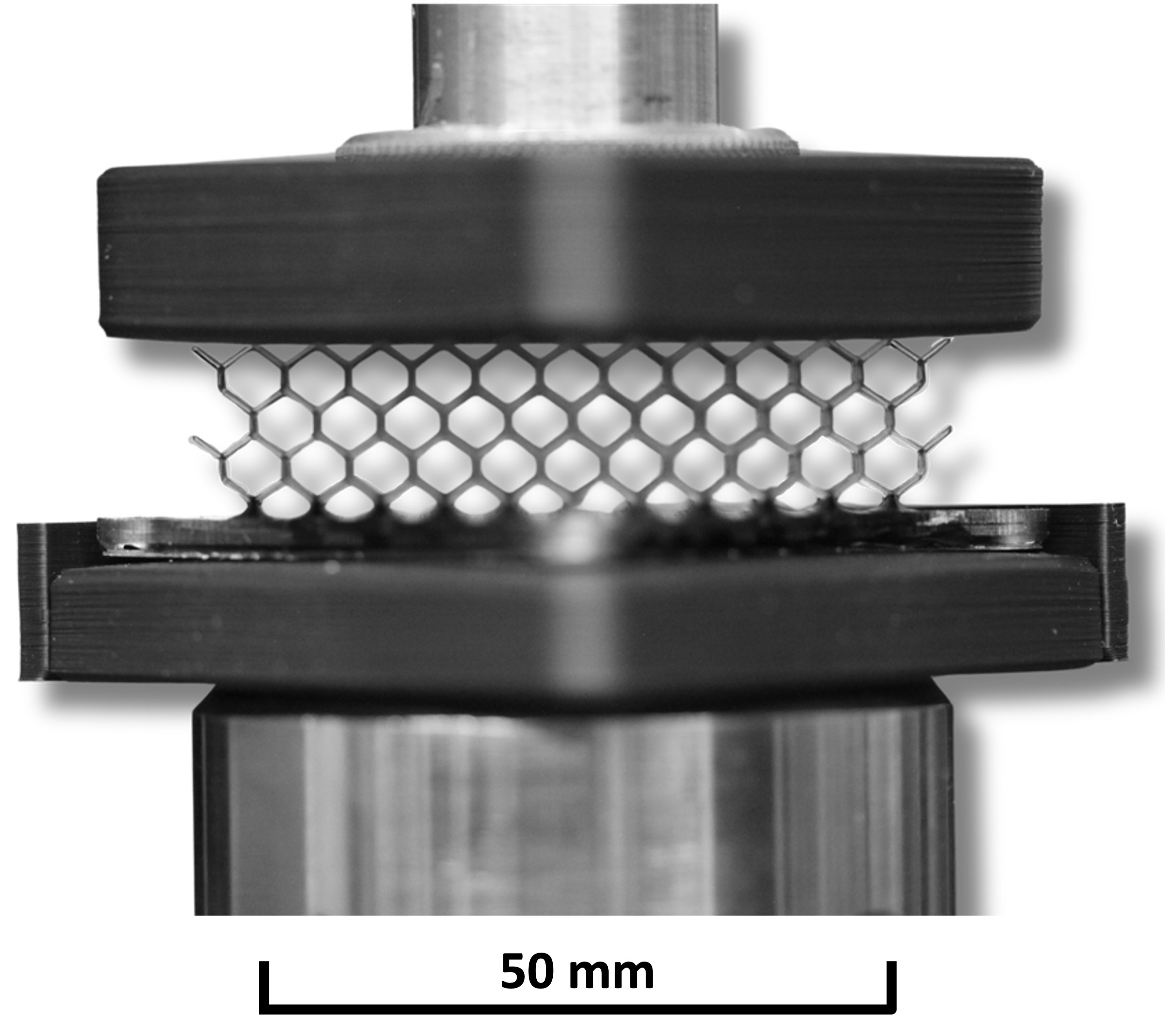}
\caption{Compression test device with diamond (D) sample}
\label{fig:photo_compression}
\end {center}
\end{figure}

Compression tests were conducted on an Alliance RF/200 machine from the MTS System. It is equipped with a load cell with a rated load of 1000 N and a sensitivity of 2.51 mV/V. The displacement speed was set at 5 mm/min for rigid samples and 10 mm/min for flexible samples that correspond relatively at $5.60~10^{-3} s^{-1}$ and $11.1~10^{-3} s^{-1}$ strain rates. Each sample was placed on a support adapted to the sole size to center the compression (Fig.~\ref{fig:photo_compression}).

\subsection{Custom impact bench}\label{sec:bench}

For low-velocity impacts, the recommended impact standards from the literature are Charpy or Izod tests for material~\cite{da_s_vieira_comparative_2018, patterson_izod_2021} and drop-weight tower for structures~\cite{mines_drop_2013, abrate_computed_2018}. However, these tests do not accurately reflect the reality of UAS impacts, as they do not account for the possibility of the structure rebounding upon impact.

\begin{figure}[H]
\begin {center}
\includegraphics[width=0.8\textwidth]{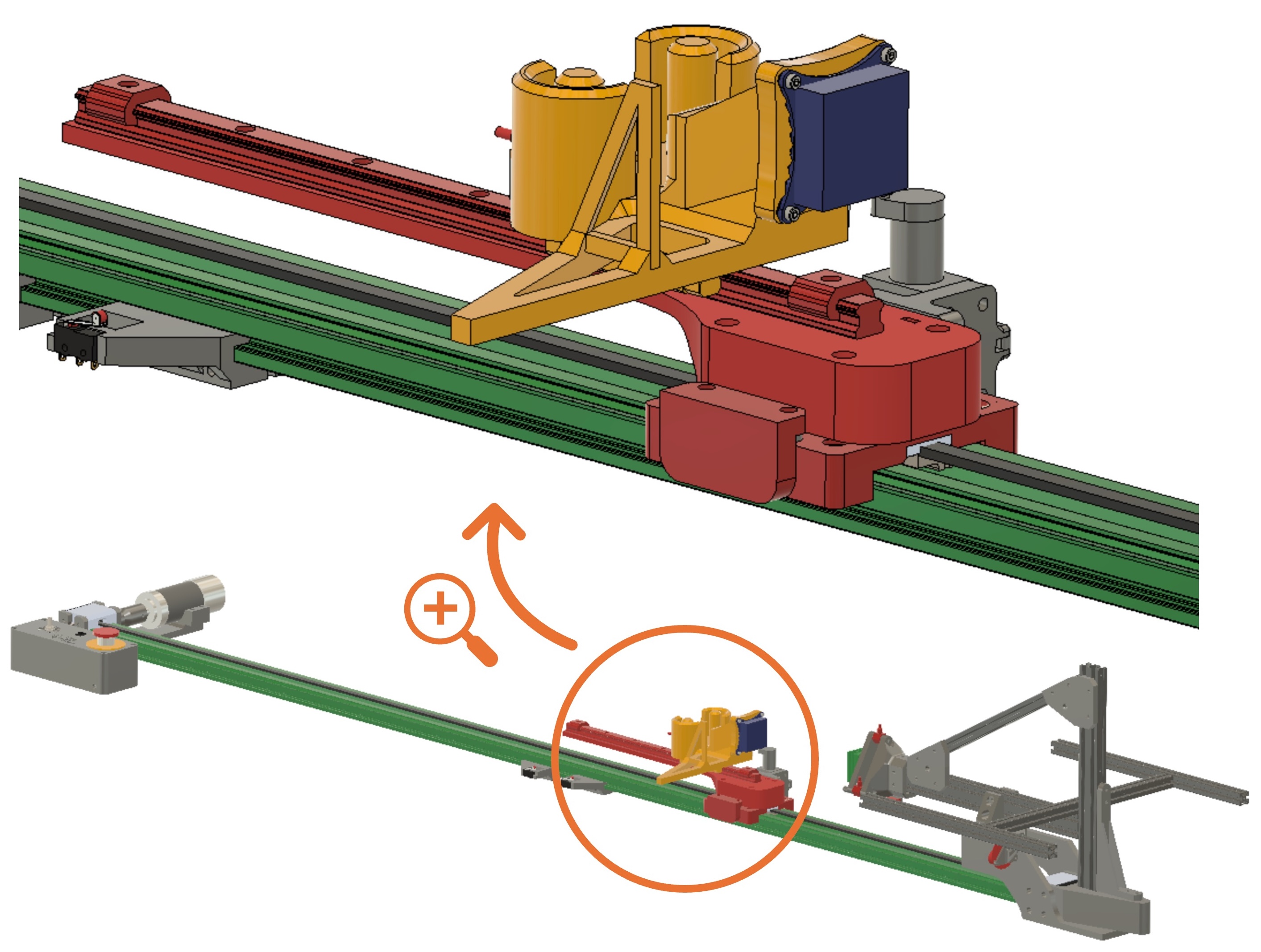}
\caption{Overview of the custom impact test bench. The first main rail (green) is used to propel the second rail (red). The sample trolley (yellow) holds the sample (blue). Just before impact, it is released to translate freely along the red rail. }
\label{fig:bench}
\end {center}
\end{figure}

We designed a custom impact test bench for this study, shown in Fig.~\ref{fig:bench}. It consists of a linear catapult driven by an electric motor, allowing for the testing of aerial vehicle structures or complete micro-UAS under conditions close to their real-world use. The bench enables controlled-speed impacts. At launch, a propulsion trolley is accelerated on the main rail. Just before impact, a sample-holding trolley, initially fixed to the propulsion trolley, is released on its secondary rail (atop the propulsion trolley). This design allows the sample to freely rebound upon impact. In this study, all impacts are performed at 2 m/s. This is representative of an inspection aerial system, where the goal is to control the position rather than the velocity, as one would do with movement phases. The trolley holding the samples is loaded with calibrated weights to simulate the inertia of the final flying device. For this work, the UAV's inertia is set to 1 kg.

The bench is equipped with a set of sensors to capture the critical and fleeting moments of the crash. Specifically, we have a PCB Piezotronics\copyright~single-axis accelerometer (352C34) to measure deceleration, three PCB Piezotronics\copyright~load cells (208C03) positioned at the vertices of an equilateral triangle to measure force during impact, a LeddarOne distance sensor from which we calculate speed before impact, and a P3 America\copyright~potentiometer (LMCR13) for high-precision distance measurement during impact. Additionally, we use a high-speed Chronos 2.0 camera to better interpret and discuss the impact behavior.

The PCB Piezotronics\copyright~sensors (accelerometer and load cells) and the potentiometer were connected to a Siemens\copyright~SCADAS acquisition system, capturing data at 6250 Hz. The Leddar sensor's data were acquired at a rate of 100 Hz using a Python script on a Linux laptop. To synchronize the acquisition systems before launch, a 5V 'trigger' signal is sent from the laptop to the SCADAS system. 

\section{Results and discussion}\label{sec:res}

Based on the raw data from the compression and impact test campaign, we calculated and represented various metrics that provide a more in-depth understanding of the micro-lattices' behavior.

\subsection{Compression tests}\label{sec:compression}

The time-lapse of the compression tests of the samples is available~\href{https://git.initrobots.ca/lcatar/micro-lattices-patches-manufacturing}{here}.

\begin{figure}[H]
\begin {center}
\includegraphics[width=1\textwidth]{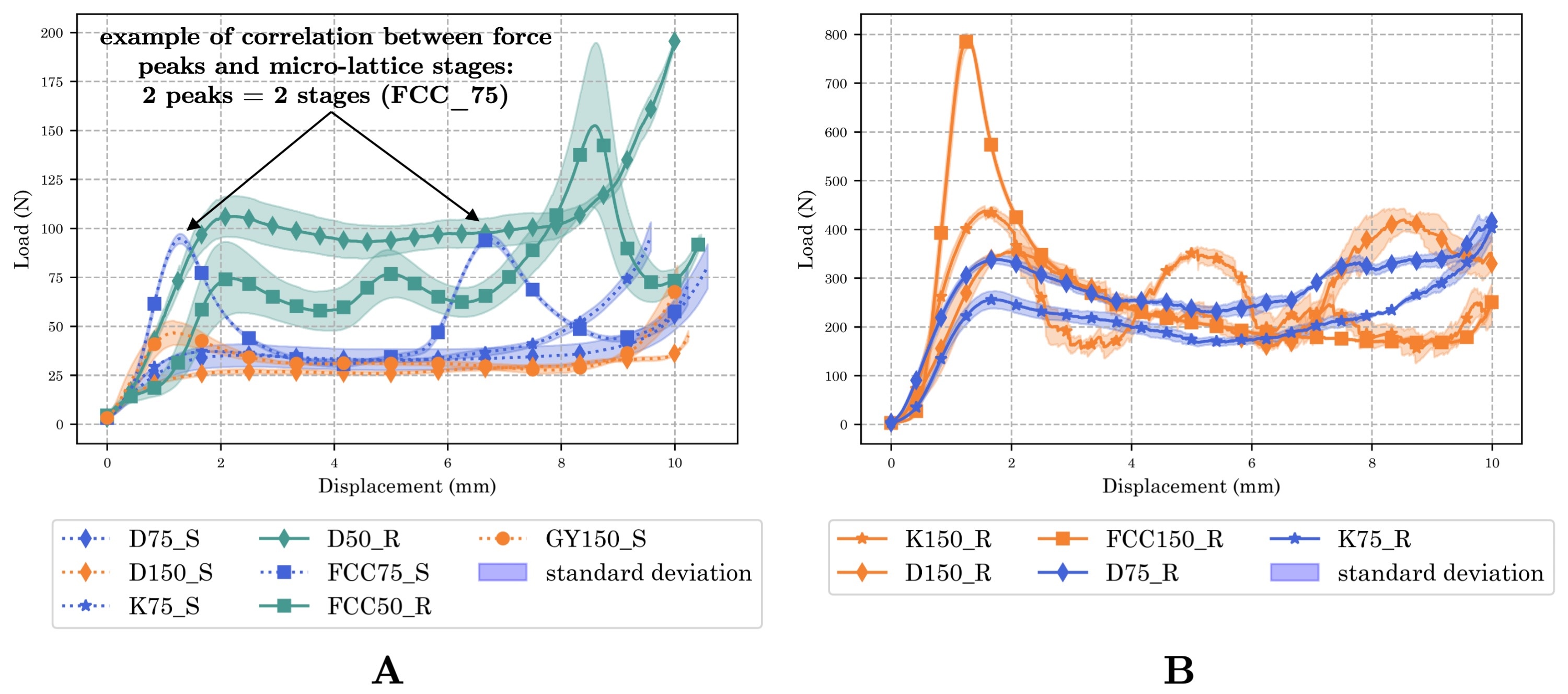}
\caption{Load-Displacement of softer (\textbf{A}) and most rigid (\textbf{B}) configurations. Both sides present similar behaviors: peaks of strength associated with loading plateaus. These peaks are particularly noticeable with the FCC pattern, directly correlated to the number of layers. Measurements' uncertainties of the three samples' repetition are represented by the colored shaded area around each curve. We deliberately did not represent the final consolidation part of the compression test to better highlight the initial variations. Therefore, the deformation stops at 10 mm and load above 100 N.}
\label{fig:most_compression}
\end {center}
\end{figure}

Several factors influence the behavior of micro-lattices under compression, which can be regrouped into geometry, compactness, and materials, as illustrated in Fig.~\ref{fig:most_compression}. Some effects can be combined. For example, the buckling of the struts in each micro-lattice depends on both the geometry and the level of compactness. This is particularly evident in the FCC micro-lattice, where the force curve shows sequential yielding stages. The number of rows per sample is directly correlated to the number of peaks in the force curve, as highlighted in (A) Fig.~\ref{fig:most_compression} with FCC\_75).

With smaller cells, we observe a longer stress plateau during loading. The same effect is seen when increasing the number of struts along the loading axis. For example, the Diamond or Kelvin patterns have more struts than the FCC pattern. The gyroid pattern, with its continuous surfaces, has a very limited contact area compared to other patterns, making the loading more variable.

In terms of material, differences in viscoelastic properties and rigidity lead to a significant decrease in maximum force, with flexible resin showing five to ten times less force than rigid resin. Larger cells (compactness 150) provide greater rigidity, even though they are fewer in number.

\begin{figure}[H]
\begin {center}
\includegraphics[width=0.8\textwidth]{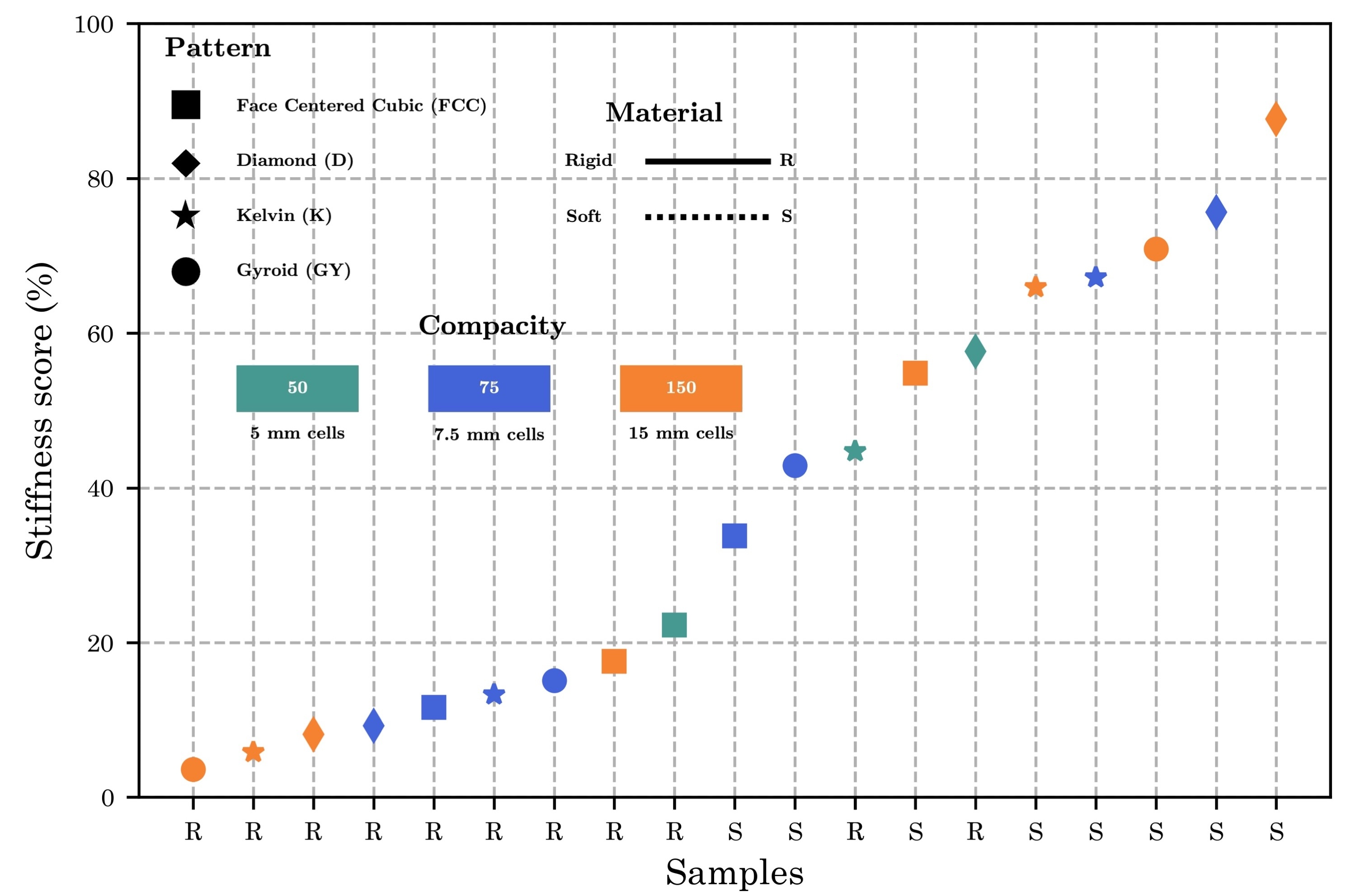}
\caption{Stiffness score for all tested combinations, showing a wide range of performances. On the right side of the graph: micro-lattices made of flexible materials obtain the highest score.}
\label{fig:stiffness}
\end {center}
\end{figure}

As previously described, each combination impacts load distribution. To emphasize this and compare all patterns, we calculated the stiffness of each micro-lattice structure. We employed an analogy with springs, defining stiffness $S_t$ as:

\begin{equation}
    S_t = \frac{d F}{d x}~[N / m]
\end{equation}
where $F$ is the force and $x$ is the displacement during the compression for each time increment. The plateau phases then strike out as their derivative is close to zero. We defined a threshold of $\pm$ 10 N/mm and each increment is labeled as a plateau or not and then reduced to a percentage of the test duration to determine a “stiffness score”. The higher the score, the more effectively the micro-lattice smooths the stress into a plateau form. Fig.~\ref{fig:stiffness} shows the stiffness score for all samples, with clear differences observed, ranging from 3.6\% to 87.7\%. 

This metric provides insight into how well each combination can adapt to constraints without creating hysteresis reactions under load. Flexible materials and diamond micro-lattices are the most effective. The order from highest to lowest stiffness score is diamond, Kelvin, and gyroid, consistently across different levels of compactness and material types. The FCC pattern shows more variability due to a regular plateau between two peaks, resulting in an excellent stiffness score between buckling stages, which nuances its overall performance. Conversely, the gyroid pattern is the least effective, likely due to the very limited contact surface between the compression surface and the micro-lattice.

\subsection{Impact tests}\label{sec:impact}

\begin{figure}[H]
\begin{center}
\includegraphics[width=1\textwidth]{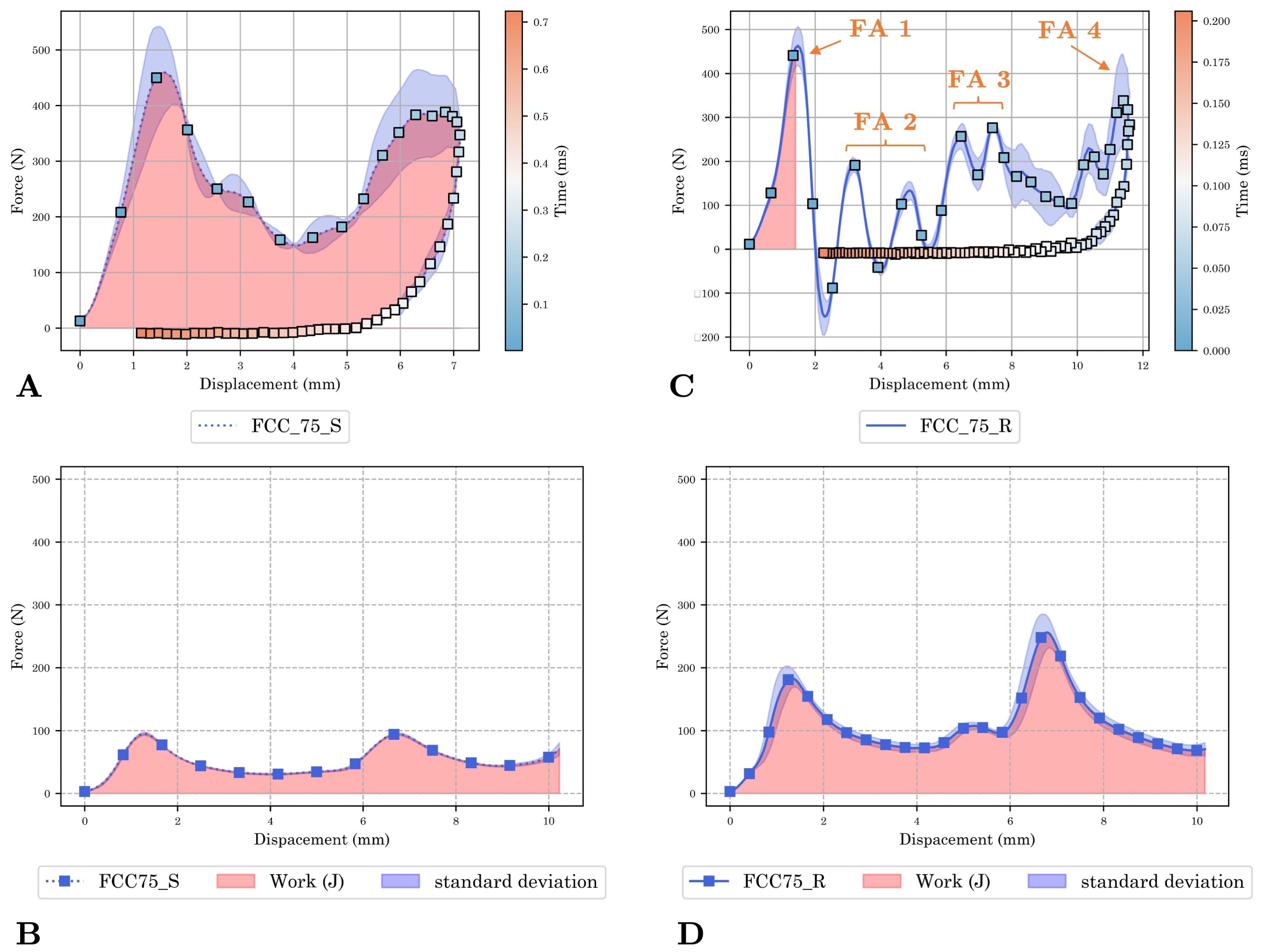}
\caption{Force with respect to displacement for the FCC micro-lattice with a compactness of 75. The curves show results for the soft material (left: A \& B) and the rigid material (right: C \& D) both for the impacts (top: A \& C) and the compression (bottom: B \& D). Under the curve, in red, the absorbed energy before the rupture of the micro-lattice is represented. Measurements' uncertainties of the three samples' repetition are represented by the colored shaded area around each curve. Different Focus Areas (FA) are highlighted in orange to support the analysis of (C).}
\label{fig:FCC_75}
\end{center}
\end{figure}

A video of all impact tests is available \href{https://git.initrobots.ca/lcatar/micro-lattices-patches-manufacturing}{here}. Fig.~\ref{fig:FCC_75} shows the typical behavior of a flexible (A) and rigid (C) micro-lattice subjected to impact. Several key observations can be made from these comparative representative results. 

First, the reaction under load is unique for each test and material. The only similarity is between the behavior of the soft and rigid micro-lattices under quasi-static compression tests (B and D). The force peaks are identical but have smaller amplitudes for the soft material. Those peaks correspond to the two layers of cells in the 75 configuration. In the soft micro-lattice under impact, these two peaks are also present, but a rebound occurs before the consolidation phase. The maximum displacement is just over 7 mm, indicating complete buckling of the first layer. However, the rebound happens before the second layer is folded.

\begin{figure}[H]
\begin{center}
\includegraphics[width=1\textwidth]{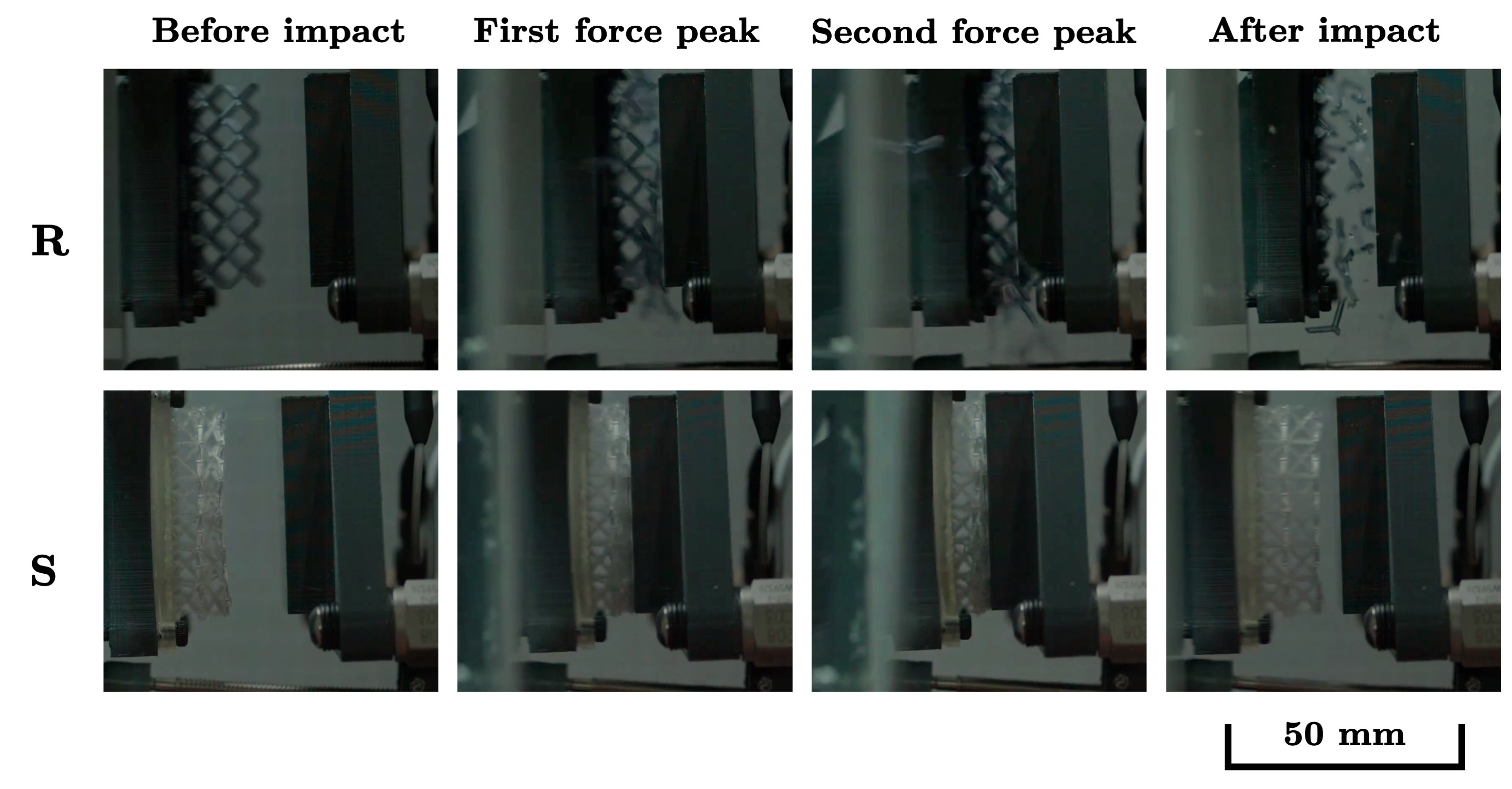}
\caption{Time-lapse of the impact of a rigid (R) and flexible (S) sample. The outcome of the impact is significantly different, with the rigid micro-lattice exploding and the flexible micro-lattice deforming without breaking. The first force peak corresponds to FA n°1 and the second force peak to FA n°3 in Fig.~\ref{fig:FCC_75}.}
\label{fig:timelapse}
\end{center}
\end{figure}

In contrast, the rigid micro-lattice shatters under impact (C), a behavior observed in all rigid samples subjected to impact. The deformation modes described by Guillon~\cite{guillon_etude_2008} and discussed in Section~\ref{sec:related} are found and verified with the high-speed camera images, of which an overview is given in Fig.~\ref{fig:timelapse}. The first peak of force corresponds to the buckling of the first stage of cells (Focus Area — FA n°1 of (C) in Fig.~\ref{fig:FCC_75}). Following the explosion of this stage, the cells were cut in two. We then observe peaks of lower amplitudes: a first series under 200 N with the bending of the struts that remained attached to the sample (FA n°2), then a second series where we go back up to nearly 300 N with the rupture of the second stage of cells located at about 7 mm of deformation (FA n°3). This distance is also similar to compression tests (rigid and flexible). The very last peak shows the shock with the sample holder trolley (FA n°4). At this stage, the micro-lattice is destroyed.

Another interesting element is that the maximum force amplitude is higher in the results of impact tests than in quasi-static compression tests. This phenomenon has been observed in the literature on numerous occasions and reported in the literature review we conducted earlier. Inertia is reported to be the main cause of this difference in absorption quality~\cite{ramakrishna_energy_1995, rathbun_performance_2006, vaughn_coupled_2005}. 

Finally, the elapsed time during the impact is represented by the color bar on the right side of the graph. The impact duration is longer for the flexible material, exceeding 0.7 s, while the rigid material reaches the same state in 0.2 s. The explosion of the micro-lattice in the rigid material involves a much more abrupt deceleration.

\subsection{Specific Energy Absorption and Efficiency}\label{sec:SEA-eta}

\begin{figure}[H]
\begin {center}
\includegraphics[width=1\textwidth]{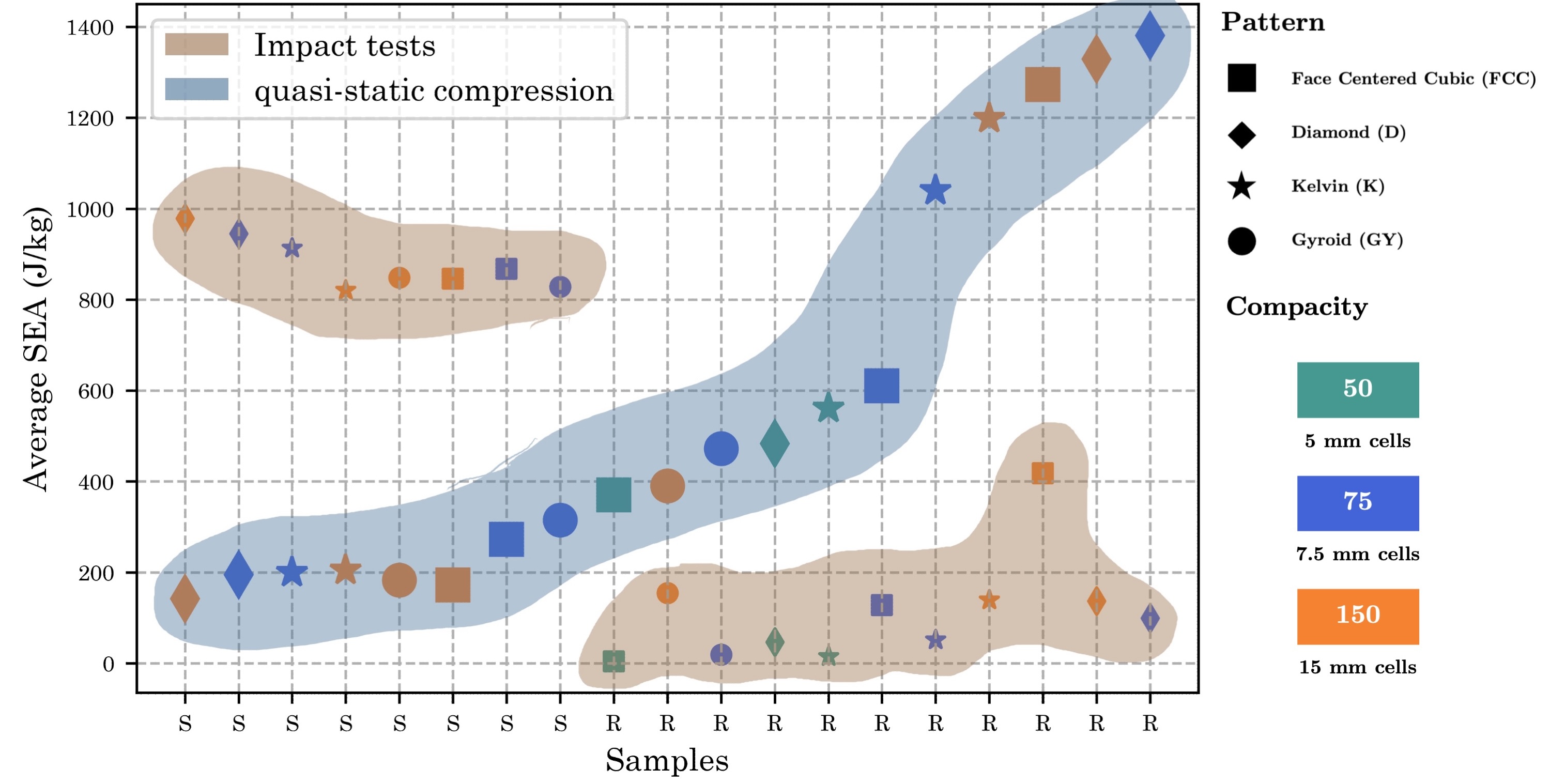}
\caption{SEA of micro-lattices specimens in quasi-static (blue shaded area) and impact (brown shaded area) tests. Rigid and soft material show an inverted tendency between the two test protocols due to the change in loading speed.}
\label{fig:SEA}
\end {center}
\end{figure}

Using Equation~\ref{eq:SEA}, we calculated the Specific Energy Absorption (SEA) of each micro-lattice combination. As noted by Ramakrisha et al.  in~\cite{ramakrishna_energy_1995}, significant variation occurs depending on the loading rate. Across all tested patterns and materials, a complete inversion between the results of quasi-static tests and impact tests is observed. Fig.~\ref{fig:SEA} shows a crossover between flexible and rigid micro-lattices: those with the highest SEA in compression exhibit the lowest SEA in impact, and vice versa.

At low loading rates, flexible micro-lattices flow due to the viscoelastic properties of the polymer, resulting in a low force below 100 N before consolidation, as seen in Fig.~\ref{fig:most_compression}~A. Rigid micro-lattices, however, undergo buckling, with the cells closing on themselves and bending while maintaining a very high reaction force throughout low-speed loading. The force experienced by the rigid micro-lattices is more than twice that of flexible ones, around 200 to 300 N, with some peaks exceeding 500 N.

In contrast, during an impact, the rigid material shatters under shock, preventing the energy from dissipating within the structure. Flexible materials, on the other hand, can withstand bending without breaking, leading to excellent performance for certain patterns at the left end of the graph.

\begin{figure}[H]
    \begin{center}
        \resizebox{1\textwidth}{!}{\input{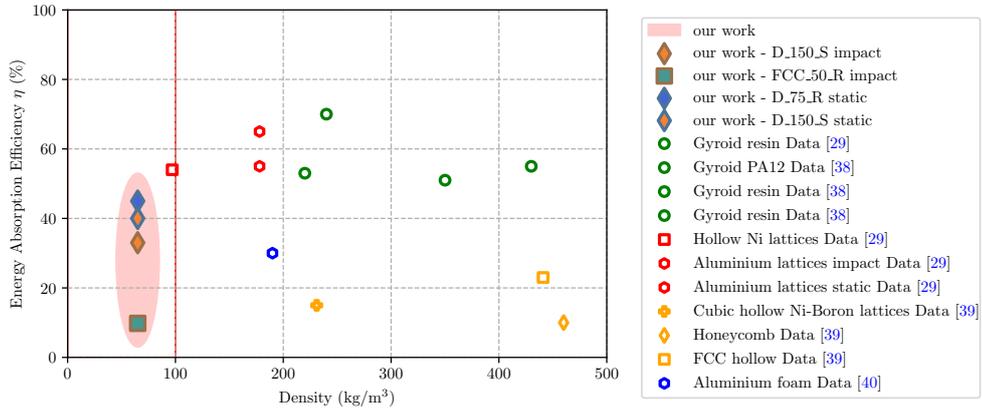}}
    \end{center}
    \caption{Energy absorption efficiency of our micro-lattices in comparison with existing works. We address the lowest density with a significant distribution of efficiency.}
    \label{fig:comparison}
\end{figure}

To compare our work with others, we plot $\eta$ with respect to the density of the micro-lattice using Equation~\ref{eq:eta}. All our samples were created with an equal mass of 2 g, resulting in a density of $65~kg/m^3$ for all specimens. In Fig.~\ref{fig:comparison}shows the micro-lattice patches with the highest and lowest $\eta$ performance from the compression and traction tests of our study. We have also included references from the literature, consisting of polymer micro-lattices made with photosensitive resins, thermoplastics manufactured using Selective Laser Sintering, and hollow metal micro-lattices.

Our work explores the limits of ultralightness, with densities set below the  $100~kg/m^3$ threshold. This ensures increased lightness with absorption-efficiency performances compared to the state-of-the-art (10\%-70\%). The micro-lattices with the closest performance in comparison are plated with nickel, which complicates implementation, especially for large-scale production. On the contrary, the use of LCD technology in additive manufacturing ensures increased ease of production and reduced production costs.

Previous work~\cite{schaedler_designing_2014}, observed less than 10\% variation in performance regardless of the configuration studied: micro-lattice, foam, or honeycomb. In contrast, our tests show an efficiency range exceeding 35\%.

Our findings are particularly valuable in cases where lightness is a critical requirement, such as the development of micro-UAS weighing less than 250 g, which falls within a more advantageous regulatory limit~\cite{hassanalian_classifications_2017}.

\subsection{Association of multiple micro-lattices}

For practical use in aerial vehicles, subjects to complex loading profiles, several micro-lattices will likely need to be combined. The differences in mechanical behavior between micro-lattices with soft or rigid resin offer a range of possibilities while maintaining the same lightness. Therefore, we studied the combination of a soft patch and a rigid patch. This combination is possible if two criteria are respected: first, the force absorbed by the soft micro-lattice must be lower than that of the rigid micro-lattice so that the soft micro-lattice deforms first; second, the energy absorbed by the soft micro-lattice must be greater than the energy required to break the rigid micro-lattice. Using the results of the impact tests, we calculated all the necessary metrics.

\begin{figure}[H]
\begin {center}
\includegraphics[width=0.45\textwidth]{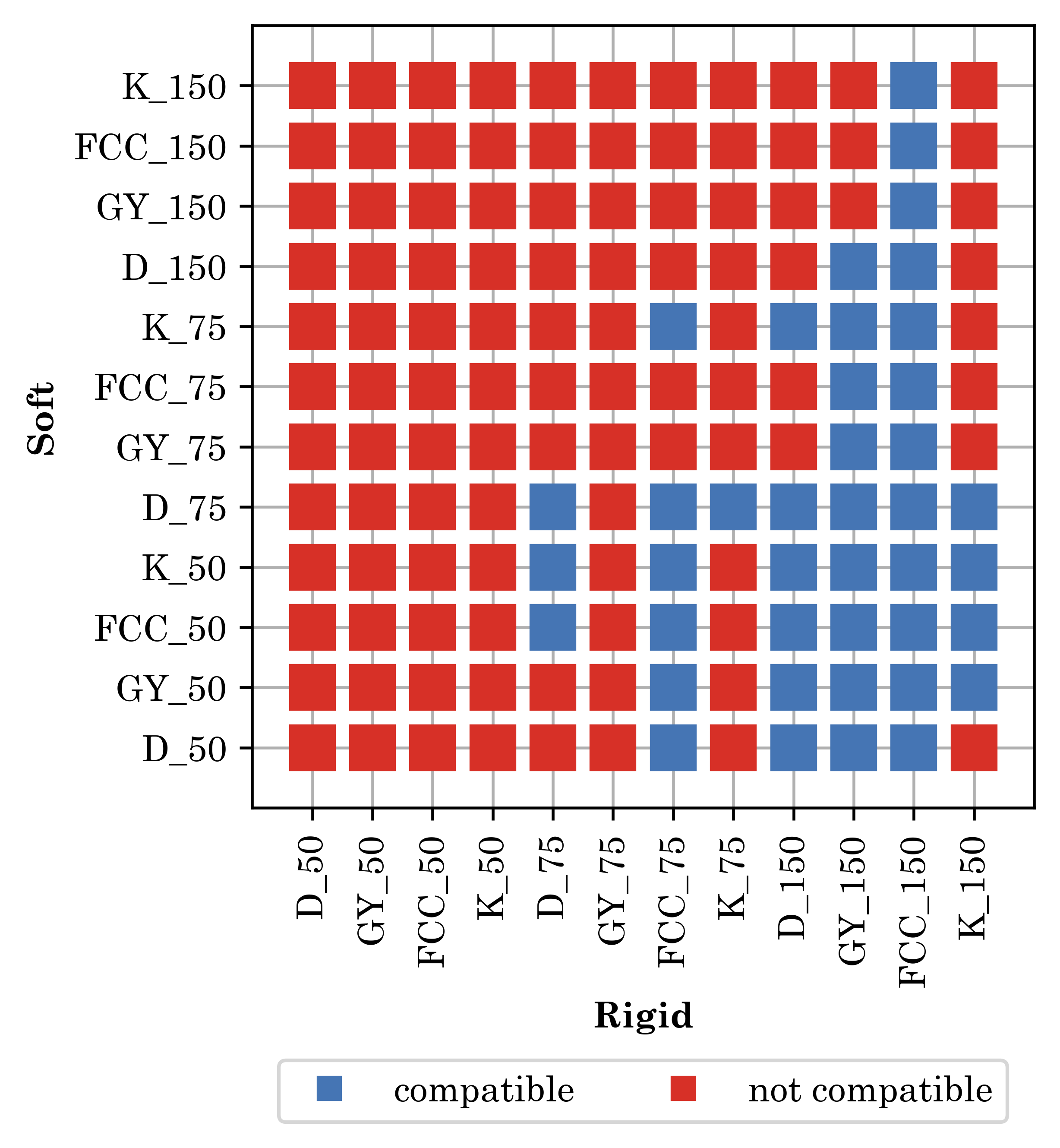}
\caption{Result of possible associations between rigid and soft micro-lattices. Most combinations with rigid material and compactness 50 or soft material and compactness 150 are eliminated by the two association criteria.}
\label{fig:association}
\end {center}
\end{figure}

From the definition of kinetic energy, we derive the speed required to break the micro-lattice. Table~\ref{tab:energy_rigid} shows the velocity values associated with each rigid combination. This “speed table” translates the energy level that must not be exceeded when using a rigid micro-lattice as a structural element in combination with a soft micro-lattice. By evaluating all possible soft and rigid combinations, we can create a compatibility map that identifies associations meeting the two criteria stated above in Fig.~\ref{fig:association}.

\begin{table}[h!]
\caption{Energy and velocity required to break rigid micro-lattices.}
    \centering
    \begin{tabular}{|l|c|c|}
        \hline
        \rowcolor{white}
        \textbf{Sample} & \textbf{Energy (J)} & \textbf{Max speed (m/s)} \\
        \hline
        \rowcolor{white} GY\_50\_R & none & none \\
        \rowcolor{red!30} FCC\_50\_R & 0.0084 & 0.1300 \\
        \rowcolor{red!20} K\_50\_R & 0.0298 & 0.2441 \\
        \rowcolor{orange!40} GY\_75\_R & 0.0392 & 0.279 \\
        \rowcolor{orange!30} D\_50\_R & 0.0941 & 0.4339 \\
        \rowcolor{orange!20} K\_75\_R & 0.1041 & 0.4563 \\
        \rowcolor{yellow!70} D\_75\_R & 0.1980 & 0.6295 \\
        \rowcolor{yellow!60} FCC\_75\_R & 0.2553 & 0.7146 \\
        \rowcolor{yellow!50} D\_150\_R & 0.2741 & 0.7404 \\
        \rowcolor{yellow!40} K\_150\_R & 0.2798 & 0.7481 \\
        \rowcolor{yellow!30} GY\_150\_R & 0.3095 & 0.7868 \\
        \rowcolor{green!20} FCC\_150\_R & 0.8351 & 1.2924 \\
        \hline
    \end{tabular}
    \label{tab:energy_rigid}
\end{table}

This compatibility map reveals that less than 30\% of the associations are viable. We summarize the relationship between the calculated stiffness score and the SEA of each combination. For example, the 75 compactness achieves a balance between absorption efficiency and stiffness, smoothing the constraint without exceeding a threshold where the transmitted energy could break the rigid micro-lattice.

\section{Conclusion}

Ultimately, we explored three variations of micro-lattices: pattern, compactness, and materials. To protect small, lightweight aerial systems, these structures demonstrated energy-absorption efficiency comparable to previous work while significantly reducing density to below the 100 g/m³ threshold. Our study also highlighted the critical influence of loading speed by examining the same geometries of samples subjected to both compression and impact tests, revealing an inversion of the SEA between these two test protocols.

Additionally, by establishing micro-lattice association criteria, we created a table of viable combinations of rigid and flexible structures. This enables the selection of compatible patterns for use in aerial vehicles, enhancing their structural resilience.

Our test system, developed with a custom free impact test bench, allowed us to characterize our structures in a controlled environment that closely simulates real-world conditions. While this new test bench may have potential inaccuracies compared to standardized devices, we demonstrated good repeatability. We plan to address this limitation by performing a performance characterization of our impact bench against a standard drop tower in future work.

This research not only advances the understanding of micro-lattice behavior under different loading conditions but also offers practical solutions for enhancing the durability of aerial vehicles. The significant density reduction and tailored energy absorption properties have the potential to impact the design and performance of lightweight, efficient aerial systems.

\backmatter

\section*{Declarations}

\subsection*{Funding}
The authors acknowledge the financial support of the NSERC CREATE UTILI program\footnote{https://carleton.ca/utili/}, and FRQNT Team grant (\#283381). 

\subsection*{Competing Interests}
The authors have no relevant financial or non-financial interests to disclose.

\subsection*{Author Contributions}
All authors contributed to the conception and design of the study. Louis Catar performed material preparation, data collection, and analysis. The first draft of the manuscript was written by Louis Catar and all authors commented on previous versions of the manuscript. All authors read and approved the final manuscript.

\bibliography{references}


\begin{thebibliography}{40}
\ifx \bisbn   \undefined \def \bisbn  #1{ISBN #1}\fi
\ifx \binits  \undefined \def \binits#1{#1}\fi
\ifx \bauthor  \undefined \def \bauthor#1{#1}\fi
\ifx \batitle  \undefined \def \batitle#1{#1}\fi
\ifx \bjtitle  \undefined \def \bjtitle#1{#1}\fi
\ifx \bvolume  \undefined \def \bvolume#1{\textbf{#1}}\fi
\ifx \byear  \undefined \def \byear#1{#1}\fi
\ifx \bissue  \undefined \def \bissue#1{#1}\fi
\ifx \bfpage  \undefined \def \bfpage#1{#1}\fi
\ifx \blpage  \undefined \def \blpage #1{#1}\fi
\ifx \burl  \undefined \def \burl#1{\textsf{#1}}\fi
\ifx \doiurl  \undefined \def \doiurl#1{\url{https://doi.org/#1}}\fi
\ifx \betal  \undefined \def \betal{\textit{et al.}}\fi
\ifx \binstitute  \undefined \def \binstitute#1{#1}\fi
\ifx \binstitutionaled  \undefined \def \binstitutionaled#1{#1}\fi
\ifx \bctitle  \undefined \def \bctitle#1{#1}\fi
\ifx \beditor  \undefined \def \beditor#1{#1}\fi
\ifx \bpublisher  \undefined \def \bpublisher#1{#1}\fi
\ifx \bbtitle  \undefined \def \bbtitle#1{#1}\fi
\ifx \bedition  \undefined \def \bedition#1{#1}\fi
\ifx \bseriesno  \undefined \def \bseriesno#1{#1}\fi
\ifx \blocation  \undefined \def \blocation#1{#1}\fi
\ifx \bsertitle  \undefined \def \bsertitle#1{#1}\fi
\ifx \bsnm \undefined \def \bsnm#1{#1}\fi
\ifx \bsuffix \undefined \def \bsuffix#1{#1}\fi
\ifx \bparticle \undefined \def \bparticle#1{#1}\fi
\ifx \barticle \undefined \def \barticle#1{#1}\fi
\bibcommenthead
\ifx \bconfdate \undefined \def \bconfdate #1{#1}\fi
\ifx \botherref \undefined \def \botherref #1{#1}\fi
\ifx \url \undefined \def \url#1{\textsf{#1}}\fi
\ifx \bchapter \undefined \def \bchapter#1{#1}\fi
\ifx \bbook \undefined \def \bbook#1{#1}\fi
\ifx \bcomment \undefined \def \bcomment#1{#1}\fi
\ifx \oauthor \undefined \def \oauthor#1{#1}\fi
\ifx \citeauthoryear \undefined \def \citeauthoryear#1{#1}\fi
\ifx \endbibitem  \undefined \def \endbibitem {}\fi
\ifx \bconflocation  \undefined \def \bconflocation#1{#1}\fi
\ifx \arxivurl  \undefined \def \arxivurl#1{\textsf{#1}}\fi
\csname PreBibitemsHook\endcsname

\bibitem[\protect\citeauthoryear{}{}]{noauthor_elios_nodate}
\begin{botherref}
Elios 3 - {Digitizing} the inaccessible.
\url{https://www.flyability.com/elios-3}
Accessed 2024-07-13
\end{botherref}
\endbibitem

\bibitem[\protect\citeauthoryear{Zheng et~al.}{2014}]{zheng_ultralight_2014}
\begin{barticle}
\bauthor{\bsnm{Zheng}, \binits{X.}},
\bauthor{\bsnm{Lee}, \binits{H.}},
\bauthor{\bsnm{Weisgraber}, \binits{T.H.}},
\bauthor{\bsnm{Shusteff}, \binits{M.}},
\bauthor{\bsnm{DeOtte}, \binits{J.}},
\bauthor{\bsnm{Duoss}, \binits{E.B.}},
\bauthor{\bsnm{Kuntz}, \binits{J.D.}},
\bauthor{\bsnm{Biener}, \binits{M.M.}},
\bauthor{\bsnm{Ge}, \binits{Q.}},
\bauthor{\bsnm{Jackson}, \binits{J.A.}},
\bauthor{\bsnm{Kucheyev}, \binits{S.O.}},
\bauthor{\bsnm{Fang}, \binits{N.X.}},
\bauthor{\bsnm{Spadaccini}, \binits{C.M.}}:
\batitle{Ultralight, ultrastiff mechanical metamaterials}.
\bjtitle{Science}
\bvolume{344}(\bissue{6190}),
\bfpage{1373}--\blpage{1377}
(\byear{2014})
\doiurl{10.1126/science.1252291} .
\bcomment{Publisher: American Association for the Advancement of Science
  Section: Report}
\end{barticle}
\endbibitem

\bibitem[\protect\citeauthoryear{Evans
  et~al.}{1998}]{evans_multifunctionality_1998}
\begin{barticle}
\bauthor{\bsnm{Evans}, \binits{A.G.}},
\bauthor{\bsnm{Hutchinson}, \binits{J.W.}},
\bauthor{\bsnm{Ashby}, \binits{M.F.}}:
\batitle{Multifunctionality of cellular metal systems}.
\bjtitle{Progress in Materials Science}
\bvolume{43}(\bissue{3}),
\bfpage{171}--\blpage{221}
(\byear{1998})
\doiurl{10.1016/s0079-6425(98)00004-8} .
\bcomment{MAG ID: 2121323562}
\end{barticle}
\endbibitem

\bibitem[\protect\citeauthoryear{Yeo et~al.}{2019}]{yeo_structurally_2019}
\begin{barticle}
\bauthor{\bsnm{Yeo}, \binits{S.J.}},
\bauthor{\bsnm{Oh}, \binits{M.J.}},
\bauthor{\bsnm{Yoo}, \binits{P.J.}}:
\batitle{Structurally {Controlled} {Cellular} {Architectures} for
  {High}-{Performance} {Ultra}-{Lightweight} {Materials}}.
\bjtitle{Advanced Materials}
\bvolume{31}(\bissue{34}),
\bfpage{1803670}
(\byear{2019})
\doiurl{10.1002/adma.201803670} .
\bcomment{\_eprint:
  https://onlinelibrary.wiley.com/doi/pdf/10.1002/adma.201803670}.
Accessed 2022-05-24
\end{barticle}
\endbibitem

\bibitem[\protect\citeauthoryear{Fina et~al.}{2018}]{fina_3d_2018}
\begin{barticle}
\bauthor{\bsnm{Fina}, \binits{F.}},
\bauthor{\bsnm{Goyanes}, \binits{A.}},
\bauthor{\bsnm{Madla}, \binits{C.M.}},
\bauthor{\bsnm{Awad}, \binits{A.}},
\bauthor{\bsnm{Trenfield}, \binits{S.J.}},
\bauthor{\bsnm{Kuek}, \binits{J.M.}},
\bauthor{\bsnm{Patel}, \binits{P.}},
\bauthor{\bsnm{Gaisford}, \binits{S.}},
\bauthor{\bsnm{Basit}, \binits{A.W.}}:
\batitle{{3D} printing of drug-loaded gyroid lattices using selective laser
  sintering}.
\bjtitle{International Journal of Pharmaceutics}
\bvolume{547}(\bissue{1}),
\bfpage{44}--\blpage{52}
(\byear{2018})
\doiurl{10.1016/j.ijpharm.2018.05.044} .
Accessed 2024-07-05
\end{barticle}
\endbibitem

\bibitem[\protect\citeauthoryear{Xiong et~al.}{2015}]{xiong_advanced_2015}
\begin{barticle}
\bauthor{\bsnm{Xiong}, \binits{J.}},
\bauthor{\bsnm{Mines}, \binits{R.}},
\bauthor{\bsnm{Ghosh}, \binits{R.}},
\bauthor{\bsnm{Vaziri}, \binits{A.}},
\bauthor{\bsnm{Ma}, \binits{L.}},
\bauthor{\bsnm{Ohrndorf}, \binits{A.}},
\bauthor{\bsnm{Christ}, \binits{H.-J.}},
\bauthor{\bsnm{Wu}, \binits{L.}}:
\batitle{Advanced {Micro}-{Lattice} {Materials}}.
\bjtitle{Advanced Engineering Materials}
\bvolume{17}(\bissue{9}),
\bfpage{1253}--\blpage{1264}
(\byear{2015})
\doiurl{10.1002/adem.201400471} .
\bcomment{\_eprint:
  https://onlinelibrary.wiley.com/doi/pdf/10.1002/adem.201400471}
\end{barticle}
\endbibitem

\bibitem[\protect\citeauthoryear{Wu et~al.}{2023}]{wu_additively_2023}
\begin{barticle}
\bauthor{\bsnm{Wu}, \binits{Y.}},
\bauthor{\bsnm{Fang}, \binits{J.}},
\bauthor{\bsnm{Wu}, \binits{C.}},
\bauthor{\bsnm{Li}, \binits{C.}},
\bauthor{\bsnm{Sun}, \binits{G.}},
\bauthor{\bsnm{Li}, \binits{Q.}}:
\batitle{Additively manufactured materials and structures: {A} state-of-the-art
  review on their mechanical characteristics and energy absorption}.
\bjtitle{International Journal of Mechanical Sciences}
\bvolume{246},
\bfpage{108102}
(\byear{2023})
\doiurl{10.1016/j.ijmecsci.2023.108102}
\end{barticle}
\endbibitem

\bibitem[\protect\citeauthoryear{Nazir et~al.}{2019}]{nazir_state---art_2019}
\begin{barticle}
\bauthor{\bsnm{Nazir}, \binits{A.}},
\bauthor{\bsnm{Abate}, \binits{K.M.}},
\bauthor{\bsnm{Kumar}, \binits{A.}},
\bauthor{\bsnm{Jeng}, \binits{J.-Y.}}:
\batitle{A state-of-the-art review on types, design, optimization, and additive
  manufacturing of cellular structures}.
\bjtitle{The International Journal of Advanced Manufacturing Technology}
\bvolume{104}(\bissue{9}),
\bfpage{3489}--\blpage{3510}
(\byear{2019})
\doiurl{10.1007/s00170-019-04085-3} .
\bcomment{MAG ID: 2956815218}
\end{barticle}
\endbibitem

\bibitem[\protect\citeauthoryear{Mines et~al.}{2013}]{mines_drop_2013}
\begin{barticle}
\bauthor{\bsnm{Mines}, \binits{R.A.W.}},
\bauthor{\bsnm{Tsopanos}, \binits{S.}},
\bauthor{\bsnm{Shen}, \binits{Y.}},
\bauthor{\bsnm{Hasan}, \binits{R.}},
\bauthor{\bsnm{McKown}, \binits{S.T.}}:
\batitle{Drop weight impact behaviour of sandwich panels with metallic micro
  lattice cores}.
\bjtitle{International Journal of Impact Engineering}
\bvolume{60},
\bfpage{120}--\blpage{132}
(\byear{2013})
\doiurl{10.1016/j.ijimpeng.2013.04.007}
\end{barticle}
\endbibitem

\bibitem[\protect\citeauthoryear{Blazy}{2003}]{blazy_comportement_2003}
\begin{botherref}
\oauthor{\bsnm{Blazy}, \binits{J.-S.}}:
Comportement mécanique des mousses d'aluminium: caractérisations
  expérimentales sous sollicitations complexes et simulations numériques dans
  le cadre de l'élasto-plasticité compressible.
PhD thesis
(2003)
\end{botherref}
\endbibitem

\bibitem[\protect\citeauthoryear{Clough et~al.}{2019}]{clough_elastomeric_2019}
\begin{barticle}
\bauthor{\bsnm{Clough}, \binits{E.C.}},
\bauthor{\bsnm{Plaisted}, \binits{T.A.}},
\bauthor{\bsnm{Eckel}, \binits{Z.C.}},
\bauthor{\bsnm{Cante}, \binits{K.}},
\bauthor{\bsnm{Hundley}, \binits{J.M.}},
\bauthor{\bsnm{Schaedler}, \binits{T.A.}}:
\batitle{Elastomeric {Microlattice} {Impact} {Attenuators}}.
\bjtitle{Matter}
\bvolume{1}(\bissue{6}),
\bfpage{1519}--\blpage{1531}
(\byear{2019})
\doiurl{10.1016/j.matt.2019.10.004}
\end{barticle}
\endbibitem

\bibitem[\protect\citeauthoryear{Ismail et~al.}{2019}]{ismail_low_2019}
\begin{barticle}
\bauthor{\bsnm{Ismail}, \binits{M.F.}},
\bauthor{\bsnm{Sultan}, \binits{M.T.H.}},
\bauthor{\bsnm{Hamdan}, \binits{A.}},
\bauthor{\bsnm{Shah}, \binits{A.U.M.}},
\bauthor{\bsnm{Jawaid}, \binits{M.}}:
\batitle{Low velocity impact behaviour and post-impact characteristics of
  kenaf/glass hybrid composites with various weight ratios}.
\bjtitle{Journal of Materials Research and Technology}
\bvolume{8}(\bissue{3}),
\bfpage{2662}--\blpage{2673}
(\byear{2019})
\doiurl{10.1016/j.jmrt.2019.04.005} .
Accessed 2023-04-27
\end{barticle}
\endbibitem

\bibitem[\protect\citeauthoryear{{SakshiKokil-Shah}
  et~al.}{2021}]{sakshikokil-shah_recent_2021}
\begin{barticle}
\bauthor{\bsnm{{SakshiKokil-Shah}}},
\bauthor{\bsnm{Sur}, \binits{A.}},
\bauthor{\bsnm{Darvekar}, \binits{S.}},
\bauthor{\bsnm{Shah}, \binits{M.}}:
\batitle{Recent {Advancements} of {Micro}-{Lattice} {Structures}:
  {Application}, {Manufacturing} {Methods}, {Mechanical} {Properties},
  {Topologies} and {Challenges}}.
\bjtitle{Arabian Journal for Science and Engineering}
\bvolume{46}(\bissue{12}),
\bfpage{11587}--\blpage{11600}
(\byear{2021})
\doiurl{10.1007/s13369-021-05992-y} .
Accessed 2023-04-04
\end{barticle}
\endbibitem

\bibitem[\protect\citeauthoryear{AlMahri
  et~al.}{2021}]{almahri_evaluation_2021}
\begin{barticle}
\bauthor{\bsnm{AlMahri}, \binits{S.}},
\bauthor{\bsnm{Santiago}, \binits{R.}},
\bauthor{\bsnm{Lee}, \binits{D.-W.}},
\bauthor{\bsnm{Ramos}, \binits{H.}},
\bauthor{\bsnm{Alabdouli}, \binits{H.}},
\bauthor{\bsnm{Alteneiji}, \binits{M.}},
\bauthor{\bsnm{Guan}, \binits{Z.}},
\bauthor{\bsnm{Cantwell}, \binits{W.}},
\bauthor{\bsnm{Alves}, \binits{M.}}:
\batitle{Evaluation of the dynamic response of triply periodic minimal surfaces
  subjected to high strain-rate compression}.
\bjtitle{Additive Manufacturing}
\bvolume{46},
\bfpage{102220}
(\byear{2021})
\doiurl{10.1016/j.addma.2021.102220}
\end{barticle}
\endbibitem

\bibitem[\protect\citeauthoryear{Xiao and
  Song}{2018}]{xiao_additively-manufactured_2018}
\begin{barticle}
\bauthor{\bsnm{Xiao}, \binits{L.}},
\bauthor{\bsnm{Song}, \binits{W.}}:
\batitle{Additively-manufactured functionally graded {Ti}-{6Al}-{4V} lattice
  structures with high strength under static and dynamic loading:
  {Experiments}}.
\bjtitle{International Journal of Impact Engineering}
\bvolume{111},
\bfpage{255}--\blpage{272}
(\byear{2018})
\doiurl{10.1016/j.ijimpeng.2017.09.018} .
\bcomment{MAG ID: 2763006544}
\end{barticle}
\endbibitem

\bibitem[\protect\citeauthoryear{Evans et~al.}{2010}]{evans_concepts_2010}
\begin{barticle}
\bauthor{\bsnm{Evans}, \binits{A.G.}},
\bauthor{\bsnm{He}, \binits{M.Y.}},
\bauthor{\bsnm{Deshpande}, \binits{V.S.}},
\bauthor{\bsnm{Hutchinson}, \binits{J.W.}},
\bauthor{\bsnm{Jacobsen}, \binits{A.J.}},
\bauthor{\bsnm{Carter}, \binits{W.B.}}:
\batitle{Concepts for enhanced energy absorption using hollow micro-lattices}.
\bjtitle{International Journal of Impact Engineering}
\bvolume{37}(\bissue{9}),
\bfpage{947}--\blpage{959}
(\byear{2010})
\doiurl{10.1016/j.ijimpeng.2010.03.007}
\end{barticle}
\endbibitem

\bibitem[\protect\citeauthoryear{Cui et~al.}{2012}]{cui_dynamic_2012}
\begin{barticle}
\bauthor{\bsnm{Cui}, \binits{X.}},
\bauthor{\bsnm{Zhao}, \binits{L.}},
\bauthor{\bsnm{Wang}, \binits{Z.}},
\bauthor{\bsnm{Zhao}, \binits{H.}},
\bauthor{\bsnm{Fang}, \binits{D.}}:
\batitle{Dynamic response of metallic lattice sandwich structures to impulsive
  loading}.
\bjtitle{International Journal of Impact Engineering}
\bvolume{43},
\bfpage{1}--\blpage{5}
(\byear{2012})
\doiurl{10.1016/j.ijimpeng.2011.11.004}
\end{barticle}
\endbibitem

\bibitem[\protect\citeauthoryear{Catar et~al.}{2022}]{catar_additive_2022}
\begin{bchapter}
\bauthor{\bsnm{Catar}, \binits{L.}},
\bauthor{\bsnm{Tabiai}, \binits{I.}},
\bauthor{\bsnm{St-Onge}, \binits{D.}}:
\bctitle{Additive {Manufacturing} {Strategy} for {Ultra}-{Lightweight} {High}
  {Value}-{Added} {Components}}.
\bpublisher{American Society of Mechanical Engineers Digital Collection},
  \blocation{???}
(\byear{2022}).
\doiurl{10.1115/DETC2022-88458} .
\burl{https://dx.doi.org/10.1115/DETC2022-88458}
Accessed 2024-02-16
\end{bchapter}
\endbibitem

\bibitem[\protect\citeauthoryear{Hassan et~al.}{2023}]{hassan_design_2023}
\begin{barticle}
\bauthor{\bsnm{Hassan}, \binits{I.M.}},
\bauthor{\bsnm{Enab}, \binits{T.A.}},
\bauthor{\bsnm{Fouda}, \binits{N.}},
\bauthor{\bsnm{Eldesouky}, \binits{I.}}:
\batitle{Design, fabrication, and evaluation of functionally graded triply
  periodic minimal surface structures fabricated by {3D} printing}.
\bjtitle{Journal of the Brazilian Society of Mechanical Sciences and
  Engineering}
\bvolume{45}(\bissue{1}),
\bfpage{66}
(\byear{2023})
\doiurl{10.1007/s40430-022-03972-3}
\end{barticle}
\endbibitem

\bibitem[\protect\citeauthoryear{Guillon}{2008}]{guillon_etude_2008}
\begin{botherref}
\oauthor{\bsnm{Guillon}, \binits{D.}}:
Étude des mécanismes d’absorption d’énergie lors de l'écrasement
  progressif de structures composites à base de fibre de carbone.
{PhD} {Thesis}
(December 2008)
\end{botherref}
\endbibitem

\bibitem[\protect\citeauthoryear{Menard and Menard}{2020}]{menard_dynamic_2020}
\begin{bbook}
\bauthor{\bsnm{Menard}, \binits{K.P.}},
\bauthor{\bsnm{Menard}, \binits{N.}}:
\bbtitle{Dynamic {Mechanical} {Analysis}},
\bedition{3}rd edn.
\bpublisher{CRC Press},
\blocation{Boca Raton}
(\byear{2020}).
\doiurl{10.1201/9780429190308}
\end{bbook}
\endbibitem

\bibitem[\protect\citeauthoryear{Ramakrishna
  et~al.}{1995}]{ramakrishna_energy_1995}
\begin{barticle}
\bauthor{\bsnm{Ramakrishna}, \binits{S.}},
\bauthor{\bsnm{Hamada}, \binits{H.}},
\bauthor{\bsnm{Maekawa}, \binits{Z.}},
\bauthor{\bsnm{Sato}, \binits{H.}}:
\batitle{Energy {Absorption} {Behavior} of {Carbon}-{Fiber}-{Reinforced}
  {Thermoplastic} {Composite} {Tubes}}.
\bjtitle{Journal of Thermoplastic Composite Materials}
\bvolume{8}(\bissue{3}),
\bfpage{323}--\blpage{344}
(\byear{1995})
\doiurl{10.1177/089270579500800307} .
\bcomment{Publisher: SAGE Publications Ltd STM}
\end{barticle}
\endbibitem

\bibitem[\protect\citeauthoryear{Gümrük and
  Mines}{2013}]{gumruk_compressive_2013}
\begin{barticle}
\bauthor{\bsnm{Gümrük}, \binits{R.}},
\bauthor{\bsnm{Mines}, \binits{R.A.W.}}:
\batitle{Compressive behaviour of stainless steel micro-lattice structures}.
\bjtitle{International Journal of Mechanical Sciences}
\bvolume{68}(\bissue{68}),
\bfpage{125}--\blpage{139}
(\byear{2013})
\doiurl{10.1016/j.ijmecsci.2013.01.006} .
\bcomment{MAG ID: 2075516086}
\end{barticle}
\endbibitem

\bibitem[\protect\citeauthoryear{Rathbun
  et~al.}{2006}]{rathbun_performance_2006}
\begin{barticle}
\bauthor{\bsnm{Rathbun}, \binits{H.J.}},
\bauthor{\bsnm{Radford}, \binits{D.D.}},
\bauthor{\bsnm{Xue}, \binits{Z.}},
\bauthor{\bsnm{He}, \binits{M.Y.}},
\bauthor{\bsnm{Yang}, \binits{J.}},
\bauthor{\bsnm{Deshpande}, \binits{V.}},
\bauthor{\bsnm{Fleck}, \binits{N.A.}},
\bauthor{\bsnm{Hutchinson}, \binits{J.W.}},
\bauthor{\bsnm{Zok}, \binits{F.W.}},
\bauthor{\bsnm{Evans}, \binits{A.G.}}:
\batitle{Performance of metallic honeycomb-core sandwich beams under shock
  loading}.
\bjtitle{International Journal of Solids and Structures}
\bvolume{43}(\bissue{6}),
\bfpage{1746}--\blpage{1763}
(\byear{2006})
\doiurl{10.1016/j.ijsolstr.2005.06.079} .
\bcomment{MAG ID: 2107419865}
\end{barticle}
\endbibitem

\bibitem[\protect\citeauthoryear{Zhang et~al.}{2021}]{zhang_energy_2021}
\begin{botherref}
\oauthor{\bsnm{Zhang}, \binits{H.}},
\oauthor{\bsnm{Zhou}, \binits{H.}},
\oauthor{\bsnm{Zhou}, \binits{Z.}},
\oauthor{\bsnm{Huizhong}, \binits{Z.}},
\oauthor{\bsnm{{Zhang Xiaoyu}}},
\oauthor{\bsnm{Zhang}, \binits{X.}},
\oauthor{\bsnm{Yang}, \binits{J.}},
\oauthor{\bsnm{{Lei Hongshuai}}},
\oauthor{\bsnm{Lei}, \binits{H.}},
\oauthor{\bsnm{Han}, \binits{F.}}:
Energy absorption diagram characteristic of metallic self-supporting {3D}
  lattices fabricated by additive manufacturing and design method of energy
  absorption structure.
International Journal of Solids and Structures,
111082
(2021)
\doiurl{10.1016/j.ijsolstr.2021.111082} .
MAG ID: 3160303303
\end{botherref}
\endbibitem

\bibitem[\protect\citeauthoryear{Smith et~al.}{2013}]{smith_finite_2013}
\begin{barticle}
\bauthor{\bsnm{Smith}, \binits{M.}},
\bauthor{\bsnm{Guan}, \binits{Z.}},
\bauthor{\bsnm{Cantwell}, \binits{W.J.}}:
\batitle{Finite element modelling of the compressive response of lattice
  structures manufactured using the selective laser melting technique}.
\bjtitle{International Journal of Mechanical Sciences}
\bvolume{67},
\bfpage{28}--\blpage{41}
(\byear{2013})
\doiurl{10.1016/j.ijmecsci.2012.12.004} .
\bcomment{MAG ID: 2016460082}
\end{barticle}
\endbibitem

\bibitem[\protect\citeauthoryear{Feng et~al.}{2022}]{feng_mechanical_2022}
\begin{barticle}
\bauthor{\bsnm{Feng}, \binits{G.-z.}},
\bauthor{\bsnm{Wang}, \binits{J.}},
\bauthor{\bsnm{Li}, \binits{X.-y.}},
\bauthor{\bsnm{Xiao}, \binits{L.-j.}},
\bauthor{\bsnm{Song}, \binits{W.-d.}}:
\batitle{Mechanical behavior of {Ti}–{6Al}–{4V} lattice-walled tubes under
  uniaxial compression}.
\bjtitle{Defence Technology}
\bvolume{18}(\bissue{7}),
\bfpage{1124}--\blpage{1138}
(\byear{2022})
\doiurl{10.1016/j.dt.2021.05.012} .
Accessed 2023-02-27
\end{barticle}
\endbibitem

\bibitem[\protect\citeauthoryear{Li et~al.}{2006}]{li_compressive_2006}
\begin{barticle}
\bauthor{\bsnm{Li}, \binits{Q.M.}},
\bauthor{\bsnm{Magkiriadis}, \binits{I.}},
\bauthor{\bsnm{Harrigan}, \binits{J.J.}}:
\batitle{Compressive {Strain} at the {Onset} of {Densification} of {Cellular}
  {Solids}}.
\bjtitle{Journal of Cellular Plastics}
\bvolume{42}(\bissue{5}),
\bfpage{371}--\blpage{392}
(\byear{2006})
\doiurl{10.1177/0021955X06063519} .
\bcomment{Publisher: SAGE Publications Ltd STM}.
Accessed 2023-11-28
\end{barticle}
\endbibitem

\bibitem[\protect\citeauthoryear{Schaedler
  et~al.}{2014}]{schaedler_designing_2014}
\begin{barticle}
\bauthor{\bsnm{Schaedler}, \binits{T.A.}},
\bauthor{\bsnm{Ro}, \binits{C.J.}},
\bauthor{\bsnm{Sorensen}, \binits{A.E.}},
\bauthor{\bsnm{Eckel}, \binits{Z.}},
\bauthor{\bsnm{Yang}, \binits{S.S.}},
\bauthor{\bsnm{Carter}, \binits{W.B.}},
\bauthor{\bsnm{Jacobsen}, \binits{A.J.}}:
\batitle{Designing {Metallic} {Microlattices} for {Energy} {Absorber}
  {Applications}}.
\bjtitle{Advanced Engineering Materials}
\bvolume{16}(\bissue{3}),
\bfpage{276}--\blpage{283}
(\byear{2014})
\doiurl{10.1002/adem.201300206} .
\bcomment{\_eprint:
  https://onlinelibrary.wiley.com/doi/pdf/10.1002/adem.201300206}.
Accessed 2022-04-08
\end{barticle}
\endbibitem

\bibitem[\protect\citeauthoryear{CATAR et~al.}{2022}]{catar_micro-tensile_2022}
\begin{botherref}
\oauthor{\bsnm{CATAR}, \binits{L.}},
\oauthor{\bsnm{Tabiai}, \binits{I.}},
\oauthor{\bsnm{St-Onge}, \binits{D.}}:
Micro-tensile test machine and setup
(2022)
\doiurl{10.6084/m9.figshare.19492121.v1}
\end{botherref}
\endbibitem

\bibitem[\protect\citeauthoryear{Saremian
  et~al.}{2021}]{saremian_experimental_2021}
\begin{barticle}
\bauthor{\bsnm{Saremian}, \binits{R.}},
\bauthor{\bsnm{Badrossamay}, \binits{M.}},
\bauthor{\bsnm{Foroozmehr}, \binits{E.}},
\bauthor{\bsnm{Kadkhodaei}, \binits{M.}},
\bauthor{\bsnm{Forooghi}, \binits{F.}}:
\batitle{Experimental and numerical investigation on lattice structures
  fabricated by selective laser melting process under quasi-static and dynamic
  loadings}.
\bjtitle{The International Journal of Advanced Manufacturing Technology}
\bvolume{112}(\bissue{9}),
\bfpage{2815}--\blpage{2836}
(\byear{2021})
\doiurl{10.1007/s00170-020-06112-0} .
Accessed 2024-06-20
\end{barticle}
\endbibitem

\bibitem[\protect\citeauthoryear{Abueidda
  et~al.}{2019}]{abueidda_mechanical_2019}
\begin{barticle}
\bauthor{\bsnm{Abueidda}, \binits{D.W.}},
\bauthor{\bsnm{Elhebeary}, \binits{M.}},
\bauthor{\bsnm{Shiang}, \binits{C.-S.A.}},
\bauthor{\bsnm{Pang}, \binits{S.}},
\bauthor{\bsnm{Abu~Al-Rub}, \binits{R.K.}},
\bauthor{\bsnm{Jasiuk}, \binits{I.M.}}:
\batitle{Mechanical properties of {3D} printed polymeric {Gyroid} cellular
  structures: {Experimental} and finite element study}.
\bjtitle{Materials \& Design}
\bvolume{165},
\bfpage{107597}
(\byear{2019})
\doiurl{10.1016/j.matdes.2019.107597} .
Accessed 2024-07-06
\end{barticle}
\endbibitem

\bibitem[\protect\citeauthoryear{da~S.~Vieira
  et~al.}{2018}]{da_s_vieira_comparative_2018}
\begin{bchapter}
\bauthor{\bsnm{S.~Vieira}, \binits{J.}},
\bauthor{\bsnm{Lopes}, \binits{F.P.D.}},
\bauthor{\bsnm{Moraes}, \binits{Y.M.}},
\bauthor{\bsnm{Monteiro}, \binits{S.N.}},
\bauthor{\bsnm{M.~Margem}, \binits{F.}},
\bauthor{\bsnm{Margem}, \binits{J.I.}},
\bauthor{\bsnm{Souza}, \binits{D.}}:
\bctitle{Comparative {Mechanical} {Analysis} of {Epoxy} {Composite}
  {Reinforced} with {Malva}/{Jute} {Hybrid} {Fabric} by {Izod} and {Charpy}
  {Impact} {Test}}.
In: \beditor{\bsnm{Li}, \binits{B.}},
\beditor{\bsnm{Li}, \binits{J.}},
\beditor{\bsnm{Ikhmayies}, \binits{S.}},
\beditor{\bsnm{Zhang}, \binits{M.}},
\beditor{\bsnm{Kalay}, \binits{Y.E.}},
\beditor{\bsnm{Carpenter}, \binits{J.S.}},
\beditor{\bsnm{Hwang}, \binits{J.-Y.}},
\beditor{\bsnm{Monteiro}, \binits{S.N.}},
\beditor{\bsnm{Firrao}, \binits{D.}},
\beditor{\bsnm{Brown}, \binits{A.}},
\beditor{\bsnm{Bai}, \binits{C.}},
\beditor{\bsnm{Peng}, \binits{Z.}},
\beditor{\bsnm{Escobedo-Diaz}, \binits{J.P.}},
\beditor{\bsnm{Goswami}, \binits{R.}},
\beditor{\bsnm{Kim}, \binits{J.}} (eds.)
\bbtitle{Characterization of {Minerals}, {Metals}, and {Materials} 2018},
pp. \bfpage{177}--\blpage{183}.
\bpublisher{Springer},
\blocation{Cham}
(\byear{2018}).
\doiurl{10.1007/978-3-319-72484-3_19}
\end{bchapter}
\endbibitem

\bibitem[\protect\citeauthoryear{Patterson et~al.}{2021}]{patterson_izod_2021}
\begin{barticle}
\bauthor{\bsnm{Patterson}, \binits{A.E.}},
\bauthor{\bsnm{Pereira}, \binits{T.R.}},
\bauthor{\bsnm{Allison}, \binits{J.T.}},
\bauthor{\bsnm{Messimer}, \binits{S.L.}}:
\batitle{{IZOD} impact properties of full-density fused deposition modeling
  polymer materials with respect to raster angle and print orientation}.
\bjtitle{Proceedings of the Institution of Mechanical Engineers, Part C:
  Journal of Mechanical Engineering Science}
\bvolume{235}(\bissue{10}),
\bfpage{1891}--\blpage{1908}
(\byear{2021})
\doiurl{10.1177/0954406219840385} .
\bcomment{Publisher: IMECHE}.
Accessed 2024-06-18
\end{barticle}
\endbibitem

\bibitem[\protect\citeauthoryear{Abrate et~al.}{2018}]{abrate_computed_2018}
\begin{barticle}
\bauthor{\bsnm{Abrate}, \binits{S.}},
\bauthor{\bsnm{Epasto}, \binits{G.}},
\bauthor{\bsnm{Kara}, \binits{E.}},
\bauthor{\bsnm{Crupi}, \binits{V.}},
\bauthor{\bsnm{Guglielmino}, \binits{E.}},
\bauthor{\bsnm{Aykul}, \binits{H.}}:
\batitle{Computed tomography analysis of impact response of lightweight
  sandwich panels with micro lattice core}.
\bjtitle{Proceedings of the Institution of Mechanical Engineers, Part C:
  Journal of Mechanical Engineering Science}
\bvolume{232}(\bissue{8}),
\bfpage{1348}--\blpage{1362}
(\byear{2018})
\doiurl{10.1177/0954406218766383} .
\bcomment{Publisher: IMECHE}.
Accessed 2024-06-18
\end{barticle}
\endbibitem

\bibitem[\protect\citeauthoryear{Vaughn et~al.}{2005}]{vaughn_coupled_2005}
\begin{barticle}
\bauthor{\bsnm{Vaughn}, \binits{D.G.}},
\bauthor{\bsnm{Canning}, \binits{J.M.}},
\bauthor{\bsnm{Hutchinson}, \binits{J.W.}}:
\batitle{Coupled {Plastic} {Wave} {Propagation} and {Column} {Buckling}}.
\bjtitle{Journal of Applied Mechanics}
\bvolume{72}(\bissue{1}),
\bfpage{139}--\blpage{146}
(\byear{2005})
\doiurl{10.1115/1.1825437} .
Accessed 2023-12-08
\end{barticle}
\endbibitem

\bibitem[\protect\citeauthoryear{Hassanalian and
  Abdelkefi}{2017}]{hassanalian_classifications_2017}
\begin{barticle}
\bauthor{\bsnm{Hassanalian}, \binits{M.}},
\bauthor{\bsnm{Abdelkefi}, \binits{A.}}:
\batitle{Classifications, applications, and design challenges of drones: {A}
  review}.
\bjtitle{Progress in Aerospace Sciences}
\bvolume{91},
\bfpage{99}--\blpage{131}
(\byear{2017})
\doiurl{10.1016/j.paerosci.2017.04.003}
\end{barticle}
\endbibitem

\bibitem[\protect\citeauthoryear{Schneider and
  Kumar}{2023}]{schneider_comparative_2023}
\begin{barticle}
\bauthor{\bsnm{Schneider}, \binits{J.}},
\bauthor{\bsnm{Kumar}, \binits{S.}}:
\batitle{Comparative performance evaluation of microarchitected lattices
  processed via {SLS}, {MJ}, and {DLP} {3D} printing methods: {Experimental}
  investigation and modelling}.
\bjtitle{Journal of Materials Research and Technology}
\bvolume{26},
\bfpage{7182}--\blpage{7198}
(\byear{2023})
\doiurl{10.1016/j.jmrt.2023.09.061} .
Accessed 2023-11-27
\end{barticle}
\endbibitem

\bibitem[\protect\citeauthoryear{Mieszala
  et~al.}{2017}]{mieszala_micromechanics_2017}
\begin{barticle}
\bauthor{\bsnm{Mieszala}, \binits{M.}},
\bauthor{\bsnm{Hasegawa}, \binits{M.}},
\bauthor{\bsnm{Guillonneau}, \binits{G.}},
\bauthor{\bsnm{Bauer}, \binits{J.}},
\bauthor{\bsnm{Raghavan}, \binits{R.}},
\bauthor{\bsnm{Frantz}, \binits{C.}},
\bauthor{\bsnm{Kraft}, \binits{O.}},
\bauthor{\bsnm{Mischler}, \binits{S.}},
\bauthor{\bsnm{Michler}, \binits{J.}},
\bauthor{\bsnm{Philippe}, \binits{L.}}:
\batitle{Micromechanics of {Amorphous} {Metal}/{Polymer} {Hybrid} {Structures}
  with {3D} {Cellular} {Architectures}: {Size} {Effects}, {Buckling}
  {Behavior}, and {Energy} {Absorption} {Capability}}.
\bjtitle{Small}
\bvolume{13}(\bissue{8}),
\bfpage{1602514}
(\byear{2017})
\doiurl{10.1002/smll.201602514} .
\bcomment{\_eprint:
  https://onlinelibrary.wiley.com/doi/pdf/10.1002/smll.201602514}.
Accessed 2023-11-27
\end{barticle}
\endbibitem

\bibitem[\protect\citeauthoryear{de~Sousa
  et~al.}{2012}]{de_sousa_assessing_2012}
\begin{barticle}
\bauthor{\bsnm{Sousa}, \binits{R.A.}},
\bauthor{\bsnm{Gonçalves}, \binits{D.}},
\bauthor{\bsnm{Coelho}, \binits{R.}},
\bauthor{\bsnm{Teixeira-Dias}, \binits{F.}}:
\batitle{Assessing the effectiveness of a natural cellular material used as
  safety padding material in motorcycle helmets}.
\bjtitle{SIMULATION}
\bvolume{88}(\bissue{5}),
\bfpage{580}--\blpage{591}
(\byear{2012})
\doiurl{10.1177/0037549711414735} .
\bcomment{Publisher: SAGE Publications Ltd STM}.
Accessed 2023-11-28
\end{barticle}
\endbibitem

\end{thebibliography}

\end{document}